\RequirePackage{fix-cm}
\documentclass{svjour3}
\smartqed
\usepackage{graphicx}
\usepackage{txfonts}
\usepackage{natbib}
\usepackage{mathptmx}

\newcommand{\am}{\mathrm{am}}
\newcommand{\sn}{\mathrm{sn}}

\newcommand{\cn}{\mathrm{cn}}

\newcommand{\dn}{\mathrm{dn}}
\newcommand{\sgn}{\mathrm{sgn}}
\newcommand{\vv}[1]{\vec{#1}}

\newcommand{\rd}{\mathrm{d}}

\newcommand{\cH}{\mathcal{H}}

\begin{document}
 \title{Analytical solution of the  Colombo top problem}

\author{
 J. Haponiak
 \and
 S. Breiter
 \and
 D. Vokrouhlick\'y}

\institute{S. Breiter \at Astronomical Observatory Institute, Faculty of Physics, Adam Mickiewicz
       University, Sloneczna 36, 61-286 Poznan, Poland \\
       \email{breiter@amu.edu.pl}
       \and
J. Haponiak \at Astronomical Observatory Institute, Faculty of Physics, Adam Mickiewicz
       University, Sloneczna 36, 61-286 Poznan, Poland \\
       \email{jacek.haponiak@amu.edu.pl}
       \and
D. Vokrouhlick\'y  \at Institute of Astronomy, Charles University, Prague, V Hole\v{s}ovi\v{c}k\'ach 2,
       180 00 Prague 8, Czech Republic\\
       \email{vokrouhl@cesnet.cz}
       }

\date{Received: date / Accepted: date}

\maketitle

\begin{abstract}
The Colombo top is a basic model in the rotation dynamics of a celestial body moving on a precessing orbit and perturbed by a gravitational torque.
The paper presents a detailed study of analytical solution to this problem. By solving algebraic equations of degree 4, we provide the expressions
for the extreme points of trajectories as functions of their energy. The location of stationary points (known as the Cassini states)
is found as the function of the two parameters of the problem. Analytical solution in terms the Weierstrass and the Jacobi
elliptic functions is given for regular trajectories. Some trajectories are expressible through elementary functions: not only the homoclinic orbits,
as expected, but also a special periodic solution whose energy is equal to that of the first Cassini state (unnoticed in previous studies).


\end{abstract}

\section{Introduction}
\label{intro}

Only about $60\%$ of the Moon surface can be seen from the Earth. The first successful attempt
to explain this fact was made by \citet{Cass:1693}, who borrowed the kinematic model of the `triple Earth motion'
from \citet[][Lib. 1,Cap.~11]{Copernicus}, and applied it to the Moon with some important amendments. Retaining the postulate of the fixed angle between
the rotation axis and the orbital plane, Cassini postulated the equality of orbital period and the sidereal rotation period, which protected the far side
from being seen from the Earth. By additionally postulating that the rotation axis, the ecliptic pole, and the lunar orbit pole remain coplanar,
Cassini suppressed the possibility of revealing the complete polar caps over one lunar axis precession cycle. Two centuries later, \citet{Tiss}
rephrased these postulates as laws. The three Cassini laws of Tisserand state that: 1) rotation and orbital periods are equal, 2) the rotation axis
has a constant inclination to the ecliptic, and 3) the three axes are coplanar. The second law differs from the original Cassini's statement,
but in view of the third law, the difference is unimportant.

With the advent of the Newtonian dynamics, the question arose if the Cassini's model is consistent with equations of motion. This was a part of the
problem issued by the French  \textit{Acad\'emie Royale des Sciences} for the Prize of 1764.  In his prize dissertation and in a later work,
\citet{Lagr:1764,Lagr:1780} demonstrated that the state described by Cassini is an equilibrium of the associated differential system,
and studied small librations in its vicinity.

The work of \citet{Colo:1966AJ} brought a new understanding of the Cassini
laws and motion near the stationary configuration which they describe
in the specific case of the Moon. In particular, Colombo demonstrated that
the second and the third laws are conceptually independent from the first
law and themselves serve as a basis of an interesting dynamical problem
which describes the long-term evolution of the  spin axis of an arbitrary
rotating body. He thus dropped from his analysis
the assumption of the direct spin-orbit resonance, but kept the
assumption of general precession of the orbital plane
due to perturbations (either caused by the oblate central body, or by other
masses in the system). He showed that the long-term dynamics of the
spin axis can be described by a simple, one dimensional problem assuming
the orbital node performs a uniform precession and the inclination
remains constant. Its stationary points represent generalizations of the
Cassini second and third laws.

It was soon understood that the Colombo problem is a very suitable
starting point for analysis of the obliquity evolution of terrestrial
planets, even when the orbital node and inclination undergo more
complex evolution. A fascinating example are studies of Mars obliquity
variations in relation to this planet's past paleoclimate, starting with
\citet{Ward:1973,Ward:1974}. Tides or internal process may additionally change some
of the system's parameters, a situation relevant to all terrestrial
planet studies, including the Moon -- see \citet{Peale:1974}, \citet{Ward:1975,Ward:1982}
or \citet{WdC:1979}, to mention just few examples of early
works. Later studies of Laskar and colleagues made a masterful use of
detailed knowledge of planetary long-term dynamics and its implications
on secular evolution of their spin axes \citep[e.g.][and many other with more technical details]{LaRo:1993, LaJoRo:1993, CorLas:2001}.
Following earlier hints,
mentioned already in \citet{HarWar:1982}, applications to giant
planets were also developed in the past two decades \citep[e.g.][]{
WardHam:2004,HamWard:2004,WardCanup:2006,BLK:2009,VokNes:2015,BrasLee:2015,RogHam:2020}.

Beyond planets and satellites, studies of secular spin evolution of asteroids
flourished recently, especially after \citet{VoNeBo:2003} applied
it to explain space parallelism of spin axes of large members in the Koronis
family (see also \citet{VoNeBo:2006}). Further applications include
spin states of exoplanets \citep[e.g.][]{Atobe:2004,Atobe:2007,SaLaBo:2019},
or artificial satellites and space debris \citep[e.g][]{EfPriSi:2018}.

Taken altogether, we note that the backbone of all these studies is the
basic Colombo model. Interestingly, a systematic mathematical treatment
of this elegant Hamiltonian problem has not been significantly advanced beyond the state of art dating back
to  \citet{HenMur:1987}, and \citet{Henrard:1993}. These classical works focused
on the aspects directly related with the probability of capture into different phase space zones when a slow parameter
evolution drives the system across the homoclinic orbits, hence they paid no attention to regular orbits.
One can also recognize the Colombo top in an anonymous quadratic Hamiltonian treated by \citet{LanEli:1995}, who
focused on the qualitative study of its parametric bifurcations.

The main reason to seek for the complete analytical solution of the Colombo top is its significance
for further studies of more realistic, perturbed problems. Be it analytical perturbation techniques,
or numerical splitting methods, the knowledge of explicit time dependence of the Colombo top motion is crucial.
The present work is divided in two principal parts: Sect.~\ref{orig} explores the problem using purely geometric and algebraic
tools, whereas the integration of equations of motion is considered in Sect.~\ref{ansol}. In other words, Sect.~\ref{orig}
concerns integral curves, that become time-dependent trajectories in Sect.~\ref{ansol}.

The formulation the problem is given in Sect.~\ref{sec:eqm}, where we introduce two sets of variables: traditional $x,y,z$, and shifted
$X,Y,Z$. Throughout the text we switch between the two sets, depending on convenience. In Sect.~\ref{sec:geom} the geometric construction of the integral curves 
is shown; intersections of the curves with the $x=X=0$ meridians (their extremities in $z$) are found in Sect.
~\ref{inters}~, expressed in terms of the energy constant and of the two parameters $a,b$.
Some of these can be critical points (the Cassini states), which allows the expression in terms of the parameters only -- given in
Sect.~\ref{Cassi}. The information gained allows to partition the phase space and distinguish three types of the Colombo top problem.

In Sect.~\ref{ansol} we first provide a universal solution in terms of the Weierstrass elliptic function $\wp$ (Sect.~\ref{Weier}),
comparing various formulations of the same result. But since the Weierstrass function behaves differently in various domains of the phase space,
we reformulate the solution in terms of the more common Jacobi elliptic functions (Sect.~\ref{Jacobi:sec}).
Finally, the specific trajectories that admit solutions in terms of elementary functions are presented
in Sect.~\ref{Spec:sec}. The closing Sect.~\ref{concl} summarizes the results and their implications.

\section{The Colombo top model}
\label{orig}

\subsection{Equations of motion}
\label{sec:eqm}
The basic assumptions leading to the Colombo top problem involve a rigid body
on an orbit around some distant primary. The orbital motion might be called
Keplerian, but with one notable addition: the orbital plane rotates uniformly
around some fixed axis in the inertial space with the angular rate $\mu$.
The body is assumed to rotate
in the lowest energy state, namely about the shortest principal axis
of its inertia tensor. Following the work of Colombo (1966), we assume the
the rotation period is not in resonance with the revolution period about the
primary (see, e.g., Peale 1969 for generalizations to this situation).
Considering the quadrupole torque due to the gravitational field of the primary,
and averaging over both orbital revolution about the center and rotation cycle
about the spin axis, one easily realizes that the total angular momentum of
rotation (or, equivalently, the angular velocity $\omega$ of rotation) is
conserved. The whole dynamical problem then reduces to the analysis of the
motion of the unit vector ${\bf r}$ of rotation pole in space.

In order to introduce fundamental astronomical parameters of relevance, let
us for a moment assume $\mu=0$, i.e. the orbital plane is fixed in the inertial
space. In this simple case, ${\bf r}$ performs regular precession about the
fixed direction ${\bf K}$ of the orbital angular momentum with a frequency
\begin{equation}
 \mu_r = - \alpha \cos \varepsilon, \label{prec-reg}
\end{equation}
where $\alpha$ is the precession constant and $\varepsilon$ is the obliquity,
namely the angle between ${\bf r}$ and ${\bf K}$ ($\cos\varepsilon={\bf r}
\cdot {\bf K}$). Note the minus sign in Eq.~(\ref{prec-reg}) which indicates
polar regression in the inertial space. In this model, $\varepsilon$ stays
constant and
\begin{equation}
 \alpha = \frac{3}{2} \frac{n^2}{\omega}\frac{E_b}{(1-e^2)^{3/2}},
  \label{pc}
\end{equation}
where $n$ is the orbital mean motion about the primary, $\omega$ the angular
rotational frequency, $E_b$ the dynamical ellipticity and $e$ the orbital
eccentricity. The dynamical ellipticity expresses degree of non-sphericity of
the body and it is defined using principal moments $A\leq B\leq C$ of the inertia
tensor as
\begin{equation}
 E_b = \frac{C-(A+B)/2}{C}. \label{eb}
\end{equation}

Things become more interesting in the case where the orbital plane about the
primary is not constant. As mentioned above, the Colombo top model describes
the situation when it performs a uniform precession in the inertial space.
In particular, ${\bf K}$ revolves uniformly on a cone about a fixed direction
${\bf K}'$ in space, such that the orbital inclination $I$ with respect to
the reference plane normal to ${\bf K}'$ is constant. The magnitude of the
precession rate is $\mu$, though most often the orbital plane performs
again regression in the inertial space (assuming $I<90^\circ$). The interest
and complexity of this model revolves about a possibility
of a resonance between the two precession frequencies $-\mu$ and $\mu_r$.
In order to describe it using a simple Hamiltonian model, Colombo (1966)
observed it is useful to refer ${\bf r}$ to the reference frame
following precession of the orbital plane, thus representing ${\bf r}^\mathrm{T}
= (\sin\varepsilon\cos(h-\pi/2),\sin\varepsilon\sin(h-\pi/2),\cos\varepsilon)$.
Here, $h-\pi/2$ is a longitude reckoned from the ascending node of the orbital
plane and $\varepsilon$ is a colatitude
measured from ${\bf K}$ as above. It is actually an advantage to introduce
$H=\cos\varepsilon$ (and $\sin\varepsilon=\sqrt{1-H^2}$). This is because
in terms of the symplectic variables $(h,H)$, the Colombo top is a one degree of
freedom, conservative problem with a Hamiltonian function
\begin{equation}
 \cH_s(h,H) = -\frac{(H-b)^2}{2} - a \sqrt{1-H^2} \, \cos{h}   = E_\mathrm{H}.
\end{equation}
and two new nondimensional parameters are defined as
\begin{equation} \label{abdef}
 a = \frac{\mu}{\alpha}\,\sin I, \qquad b = \frac{\mu}{\alpha}\,\cos I.
\end{equation}

As observed by \citet{HenMur:1987}, the discussion can be confined to nonnegative constants $a$ and $b$ thanks to the symmetries $(a,h,H) \rightarrow (-a, h+\pi, H)$, and
$(b,h,H) \rightarrow (-b,-h,-H)$ admitted by $\cH_s$.

Since the equations of motion
\begin{equation}\label{eqhH}
    \dot{h} = \frac{\partial \cH_s}{\partial H} = b - H + \frac{a H \cos{h}}{\sqrt{1-H^2}},
    \qquad
    \dot{H} = - \frac{\partial \cH_s}{\partial h} = - a \sqrt{1-H^2} \sin{h},
\end{equation}
are singular at $H^2 = 1$, it is better to use the Cartesian coordinates of the unit momentum vector
\begin{equation}
    \mathbf{r}= \left(
                  \begin{array}{c}
                    x \\
                    y \\
                    z \\
                  \end{array}
                \right) =
                \left(
                  \begin{array}{c}
                    \sqrt{1-H^2} \sin{h} \\
                    -\sqrt{1-H^2} \cos{h} \\
                    H \\
                  \end{array}
                \right).
\end{equation}
Then, similarly to \citet{HenMur:1987} we obtain the Hamiltonian function
\begin{equation}\label{Hxyz}
    \cH_\mathrm{HM}(x,y,z) = - \frac{(z-b)^2}{2} + a y = E_\mathrm{H},
\end{equation}
that generates equations of motion
\begin{eqnarray}
  \dot{x} &=&  (z-b)(y+a) + a b, \nonumber \\
  \dot{y} &=& -(z-b)\,x, \label{eqm:c}\\
  \dot{z} &=& - a x. \nonumber
\end{eqnarray}

For the sake of minor simplification, let us propose the shifted variables
\begin{equation}\label{shifted}
    X=x,  \quad Y = y+a, \quad Z = z-b, \quad  \vec{R} = (X,Y,Z)^\mathrm{T} = \vec{r}+(0,a,-b)^\mathrm{T}.
\end{equation}
Adding a constant $a^2$ to the Hamiltonian (\ref{Hxyz}) we simplify it to
\begin{equation}\label{Hxyzs}
  \mathcal{H} =  - \frac{Z^2}{2} + a Y = E,
\end{equation}
with the new energy constant
\begin{equation}\label{Edef}
    E = E_\mathrm{H}+a^2.
\end{equation}
Equations of motion for the shifted variables are
\begin{eqnarray}
  \dot{X} &=& Z\,Y + a b, \nonumber \\
  \dot{Y} &=& -Z\,X, \label{eqm:C}\\
  \dot{Z} &=& - a X. \nonumber
\end{eqnarray}
Notably, when $a=0$, the problem is simplified to the symmetric free top, with
constant $Z$ and uniform rotation of $\vv{r}$ around the third axis, with the frequency
$\sqrt{- 2 E} = |Z |$.

\subsection{Geometric interpretation}

\label{sec:geom}

It is customary to represent the integral curves of Eqs.~(\ref{eqm:c}) as the intersections of two
surfaces:
\begin{description}
  \item[~~S1] -- a sphere $x^2+y^2+z^2=1$,
  \item[~~S2] -- a parabolic cylinder $(z-b)^2  - 2 a (y+a) + 2 E = 0$, implied by the energy integral.
  \end{description}
The symmetry plane of S2 is $z=b$, and its vertex line is parallel to the $x$-axis, passing through
$y= (E-a^2)/a$.

\begin{figure}
\begin{center}
 \includegraphics[width=6cm]{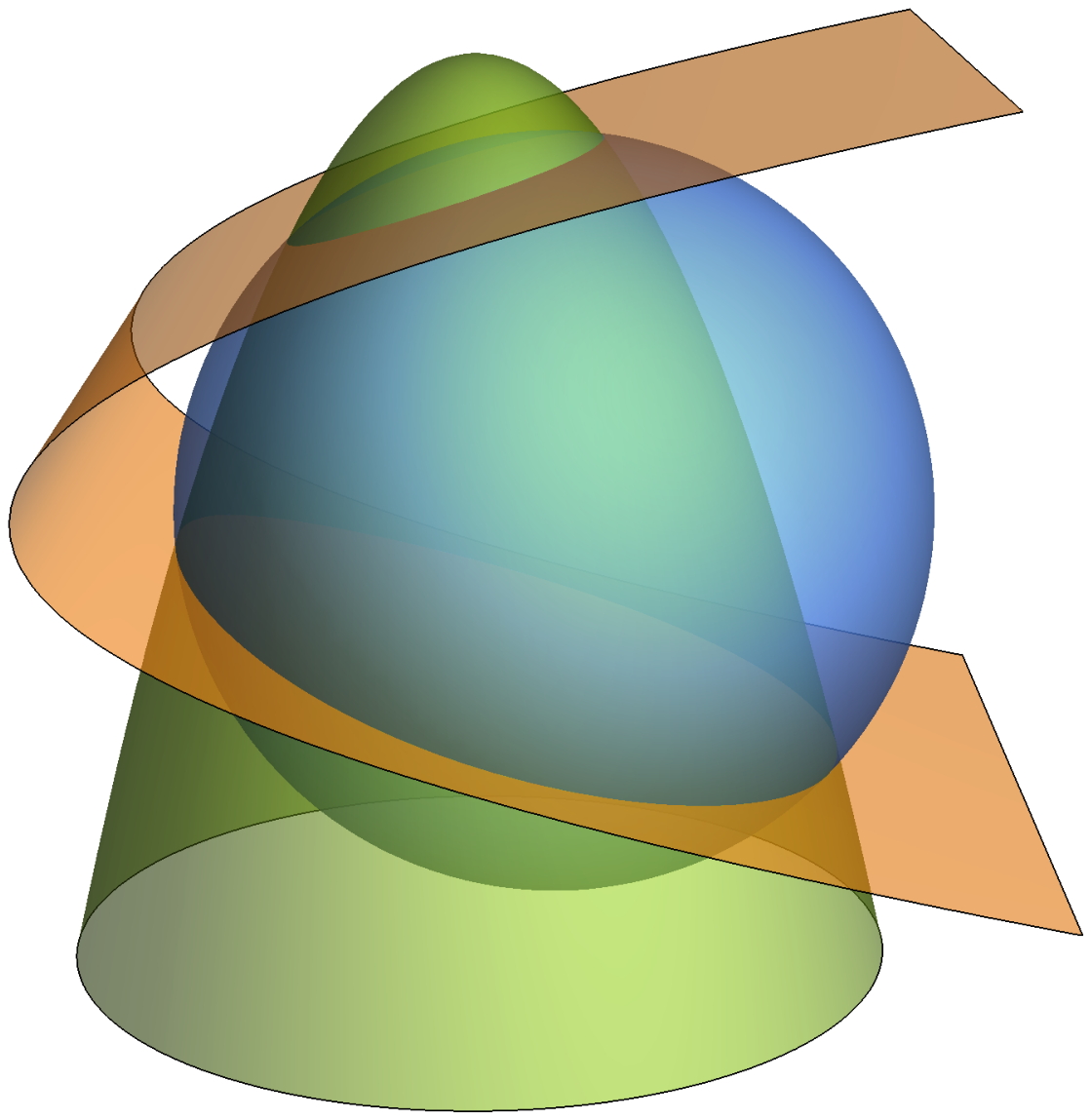}
 \includegraphics[width=6cm]{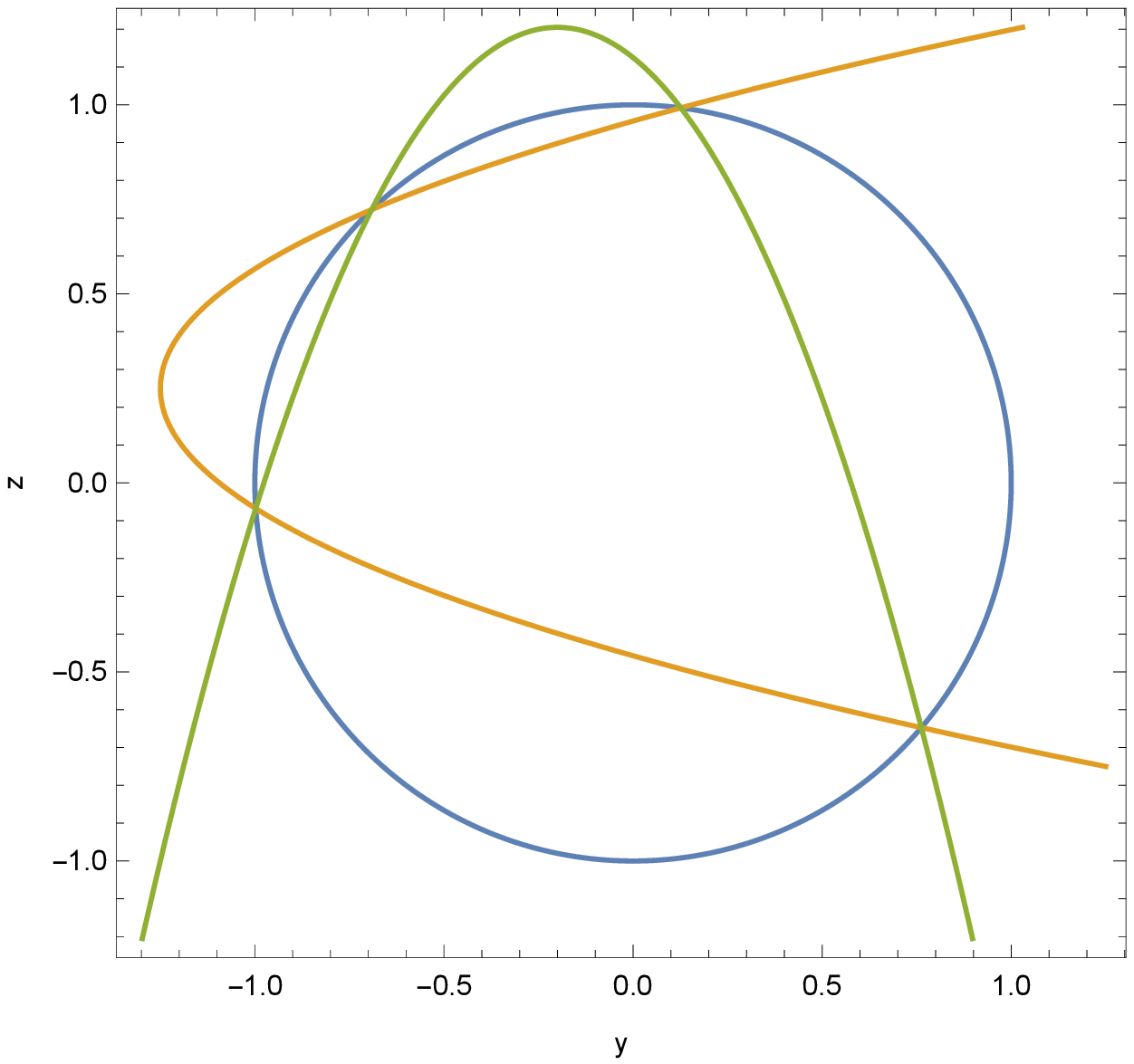}
 \end{center}
 \caption{Surfaces S1, S2, S3 (left) and their section at $x=0$ (right), for $a=0.2$, $b=0.25$, $E=-0.29$.}
 \label{fig:1}
\end{figure}

Observe that combining S1 and S2 we can also obtain another surface (see Fig.~\ref{fig:1}):
\begin{description}
  \item[~~S3] -- a paraboloid of revolution
   $x^2+(y+a)^2+ 2 b z - \left( 2E - a^2 + b^2+1 \right) = 0$.
\end{description}
Actually, the S3-based function
\begin{equation}\label{Halt}
    \cH_\mathrm{a}(\vv{r}) = \frac{x^2}{2} + \frac{(y+a)^2}{2} + b z = E' = E + \frac{1-a^2 + b^2}{2},
\end{equation}
is an alternative Hamiltonian of the Colombo top, leading to the same equations of motion (\ref{eqm:c})
as the Hamiltonian (\ref{Hxyz}).

The paraboloid S3 has the symmetry axis parallel to the $z$-axis, and passing through the points $x=0$, $y=-a$.
Its vertex is located at $z = E'/b$. The advantage of S3 appears when discussing the limit of $a=0$.
Then, the paraboloid does not change the shape and the intersections of S1 and S3 are circles because of coincidence of the
symmetry axes. Contrarily to this, setting $a=0$ in S2 results in degeneracy: the cylinder breaks in two parallel planes.
On the other hand, $b=0$ turns the $S3$ paraboloid into a (circular) cylinder, whereas S2 retains its shape.
Thus, S2 and S3 (or $\cH$ and $\cH_\mathrm{a}$) can be considered complementary, although the regularity at $a=0$
seems to us more favorable (e.g. if some perturbation approach is based upon the small inclination assumption).

An integral curve can be represented as a parametric curve in a number of ways. Of course, the best is to solve
Eqs.~(\ref{eqm:c}) and find $\vv{r}(t)$. But before we accomplish it (and, actually, in order to do it)
let us consider parameterizations, where two coordinates are expressed in terms of the third.
The most straightforward is to solve the system of S1 and S2 equations, using $z$ as a parameter variable, which leads to
\begin{equation}\label{rofz}
    \vv{r}(z) =  \left(
    \begin{array}{c}
    \pm    \sqrt{ 1-z^2  - y(z)^2}   \\
                             \frac{ (z-b)^2 + 2 E-2 a^2}{2a}     \\
                                 z \\
                                \end{array}
                              \right),
       \mbox{~~or~~}
    \vv{R}(Z) =  \left(
    \begin{array}{c}
    \pm    \sqrt{ 1-(Z+b)^2  - \left(Y(Z)-a\right)^2}   \\
                             \frac{ Z^2 + 2 E}{2a}     \\
                                 Z \\
                                \end{array}
                              \right).
\end{equation}

All three invariant surfaces S1, S2, S3 intersect the plane $x=0$, which is their common plane of symmetry.
The integral curves, as the lines of intersection of S1, S2, and S3 also pass through $x=0$ and are symmetric with respect to this plane.
Moreover, both $y$ and $z$ coordinates of a given integral curve attain their local extremes at $x=0$, according to equations of motion (\ref{eqm:C}).

\subsection{Intersections with the plane $x=0$}
\label{inters}

To find the intersection points of an integral curve with the plane $x=0$  for some specified energy $E$, it is enough to find either $y$ or $z$
coordinate. Knowing one of them, one might recover the other from the relation $y^2+z^2=1$. But this involves the ambiguity of sign, thus it is better to
find $z$, and then use the parametric equation (\ref{rofz}) for $y(z)$. To benefit from minor simplifications, we first find $Z$ and then $Y(Z)$.

According to Eq.~(\ref{rofz}), the relation between $X$ and $Z$ is
\begin{equation}\label{XofZ}
    X = \pm \frac{1}{2a}\,\sqrt{W(Z)},
\end{equation}
where $W(Z)$ is a polynomial of degree 4
\begin{equation}\label{Wgen}
    W(Z) = -  Z^4  - 4 E Z^2 - 8 a^2 b Z - 4 \left(E - a^2\right)^2 + 4 a^2 \left(1- b^2\right).
\end{equation}
Note the absence of the cubic term thanks to the use of the shifted variable $Z$.

Solving the quartic equation $W(Z)=0$ is a tedious task; the details can be found in Appendix~\ref{Ap:1}.
Briefly, the four roots $Z_j$ are given in terms of the three roots $e_i$ of the reduced cubic resolvent
equation (additionally modified to match the standard cubic  polynomial appearing in the Weierstrass elliptic integrals).
Depending on the parameters $a$ and $b$, selecting some energy value $E$, the number of meridian intersection points is 4, 3, 2, 1, or 0.

Four intersection points mean that there are two distinct integral curves with the same energy, each intersecting $x=0$ in two points:
\begin{eqnarray}
    z_1 & = & b - \sqrt{e_1-\frac{2}{3}E} - \sqrt{e_2-\frac{2}{3}E} - \sqrt{e_3-\frac{2}{3}E}, \nonumber \\
    z_2 & = & b - \sqrt{e_1-\frac{2}{3}E} + \sqrt{e_2-\frac{2}{3}E} + \sqrt{e_3-\frac{2}{3}E}, \label{z12:f}
\end{eqnarray}
for a lower curve, and
\begin{eqnarray}
    z_3 & = & b + \sqrt{e_1-\frac{2}{3}E} - \sqrt{e_2-\frac{2}{3}E} + \sqrt{e_3-\frac{2}{3}E}, \nonumber \\
    z_4 & = & b + \sqrt{e_1-\frac{2}{3}E} + \sqrt{e_2-\frac{2}{3}E} - \sqrt{e_3-\frac{2}{3}E}, \label{z34:f}
\end{eqnarray}
for the upper curve, with $e_j$ defined in Eq.~(\ref{ejBr}) as functions of $E$, $a$, and $b$,
such that $z_1 < z_2 < z_3 < z_4$.

Three intersection points mean that either one of the two curves contracts to a single point, or two curves share the same intersection point.
These situations are distinguished by the sign of invariant $g_3$ from Eq.~(\ref{g3}).
\begin{itemize}
\item If $g_3 > 0$, then the upper curve shrinks into a point with
\begin{equation}\label{zm34:f}
    z_{34} = b + \sqrt{e_1- \frac{2}{3}E}.
\end{equation}
The remaining intersection points of the lower curve are
\begin{equation}
  z_1= 2 b - z_{34} - 2 \sqrt{e_{23}-\frac{2}{3}E}, \qquad
  z_2=2 b - z_{34} + 2 \sqrt{e_{23}-\frac{2}{3}E}. \label{z12q:f}
\end{equation}
The expressions of $e_1$ and $e_{23}$ are given in Eq.~(\ref{e1e23}).
\item If $g_3 < 0$, then the upper and the lower curves meet at
\begin{equation}
    z_{23} = b + \sqrt{e_3 -\frac{2}{3}E},
\end{equation}
with the remaining intersection points at
\begin{equation}
  z_1= 2b - z_{23}  - 2 \sqrt{e_{23}-\frac{2}{3}E}, \qquad
  z_4 = 2b - z_{23}  + 2 \sqrt{e_{23}-\frac{2}{3}E}. \label{z14q:f}
\end{equation}
\end{itemize}

If $e_1=e_2=e_3$, then $z_2$ from Eq.~(\ref{z12q:f}) becomes equal to $z_{34}$ an only two intersection points remain
\begin{equation}\label{z1z234:f}
    z_1 =  b - 3 \sqrt[3]{a^2 b}, \qquad z_{234} = b+ \sqrt[3]{a^2 b},
\end{equation}
with $a^\frac{2}{3}+b^\frac{2}{3}=1$. This case requires a unique value of energy with a simple expression (\ref{Etr}).

A more generic situation with two intersection points occurs when $E$ defines only one trajectory, with $z_1 < z_2$ given by
\begin{eqnarray}
    z_1 & = & b - \sqrt{e_1-\frac{2}{3}E} - \sqrt{ e_1- 2 E + \sqrt{\nu}}, \nonumber \\
    z_2 & = & b - \sqrt{e_1-\frac{2}{3}E} + \sqrt{ e_1 - 2 E + \sqrt{\nu} }, \label{z122:f} \\
    \nu &=&  \left(e_1+ \frac{4}{3}E\right)^2+e_\mathrm{c}^2, \nonumber
\end{eqnarray}
where $e_1$ and $e_\mathrm{c}$ are defined in Eqs.~(\ref{e1:dm}) and (\ref{ec:def}). Two
other roots of $W(z-b)=0$, i.e. $z_3= Z_3+b$ and $z_4 = \overline{z}_3$ are complex, so they do not
define the intersections -- see Eqs.~(\ref{Z34:dm}).

Finally, one intersection point means that for the energy $E$ there is only one integral curve that contracted to a point with
\begin{equation}\label{zm12:f}
    z_{12} = b - \sqrt{e_1 - \frac{2}{3} E},
\end{equation}
and $e_1$ given by Eq.~(\ref{e1e23}).

In further discussion, the point of intersection of a regular curve with the $x=0$ plane will be named a turning point, unless
it is an equilibrium from the dynamical point of view.

\subsection{Cassini states and homoclinic orbits}

\label{Cassi}

Finding the Cassini states can be approached from different points of view. Geometrically, they are the points of tangency of the surfaces S1, S2, and S3.
Algebraically, they are the multiple roots of $x(z,E)=0$, and $x(y,E)=0$. Dynamically, they are the fixed points of equations of motion (\ref{eqm:c}) or,
equivalently, the local extremes and saddle points of the Hamiltonian on a sphere S1.

In principle, the multiple roots have been found in Section~\ref{inters}. But they are given in terms of $E$, which is an implicit function of $a$ and $b$
as a root of $\Delta(E,a,b)=0$, where $\Delta$ is the discriminant of $W(Z)$, defined in Eq.~(\ref{Delta}). Hence, one possible way is to find the roots of $\Delta=0$, which is
a quartic equation in $E$, and substitute them into
$z_{12}$, $z_{23}$, or $z_{34}$ from Sect.~\ref{inters} (we already did it for $z_{234}$). The alternative is to find the stationary points $(x,y,z)$ of the Hamiltonian
and use them to evaluate the energy by the substitution into $\cH(x,y,z)=E$. The latter path is more convenient and was recently taken by \citet{SaLaBo:2019}.

Discussing the Cassini states as the multiple roots of $W(Z)=0$, we should include the information, that they are also the roots of its derivative $W'(Z)=0$. Since two polynomials
have a common root if and only if their resultant equals zero, we ask about the solutions of $R(W(Z),W'(Z))=0$. If we treat $W(Z)$ and $W'(Z)$ as polynomials in $Z$,
the result is simply the discriminant $\Delta$ from Eq.~(\ref{Delta}), multiplied by a constant (negative) factor. Yet, we can also treat  $W(Z)$ and $W'(Z)$  as polynomials of $E$
with degrees 2 and 1, respectively, whose coefficients depend on $Z$.  So, using
\begin{eqnarray}
 & & W(Z)  = U_1(E) = - 4 E^2 + 4 \left( 2 a^2 - Z^2 \right) E  - Z^4 - 4 a^2 \left(2  b Z + a^2 + b^2 - 1\right), \\
 & & W'(Z)  = U_2(E) = - 8 Z E  - 4 Z^3 - 8 a^2  b,
\end{eqnarray}
we obtain the resultant  $R(U_1(E),U_2(E)) = 256 a^2 w(Z)$, where
\begin{equation}\label{wZ}
    w(Z) = - Z^4  - 2 b Z^3 + 3 \rho \, Z^2  - 2 a^2 b Z - a^2 b^2,
\end{equation}
and
\begin{equation}\label{alfa}
    \rho = \frac{1-a^2-b^2}{3}.
\end{equation}

For easier reference to \citet{SaLaBo:2019}, we abandon the shifted variables and solve $w(z-b)=0$, which is the same equation as the one
they used. However, thanks to considering the Cartesian variables on a sphere, we do not have to warn about any loss of sign, because it does not occur
neither in evaluating $R(U_1,U_2)$, nor in obtaining $W(Z)$. The value of $y$ associated with a given $z$ results directly from the condition
$\dot{x}=0$ in Eq.~(\ref{eqm:c}), being
\begin{equation} \label{yofz}
y = -a - \frac{ a b}{z-b} = - \frac{a z}{z-b}.
\end{equation}

Solving $w(z-b)=0$ we can recycle the procedure used for $W(Z)=0$. More details can be found in Appendix~{\ref{Ap:Cas}, so here we only summarize the final
results. When considering the problem on a sphere, there are only two generic situations: either there are two Cassini states $C_2$ and $C_3$, or there are
four of them: $C_1$, $C_2$, $C_3$ and $C_4$. Let us call the former case `the Colombo top problem of type II', and the latter - `type IV'. The special case
when there are three Cassini states will be called `the Colombo top problem of type III' (see Fig.~\ref{fig:2}).

In the following subsections we briefly review each type, providing the coordinates of the Cassini states $C_j$ as the functions of $a$ and $b$ parameters.
To avoid confusion with the $z_i$ expressions of Sect.~\ref{inters}, we label the coordinates of $C_j$ as $z^\ast_j$, and $y^\ast_j$ (of course,
$x^\ast_j=0$ for all the points). We only provide $z^\ast_j$ that allows to find $y^\ast_j$ from Eq.~(\ref{yofz}) and then the energy $E_j$ of the
Cassini state is
\begin{equation}\label{ECj}
    E_j = - \frac{\left(z^\ast_j-b\right)^2}{2} - \frac{ a^2 b}{z^\ast_j-b}.
\end{equation}
Substituting this energy value into an appropriate $z_{12}$, $z_{23}$, $z_{34}$, or $z_{234}$ from Sect.~\ref{inters}, should result in
returning to the expression of $z^\ast_j$ in terms of $a$ and $b$.

\begin{figure}
\begin{center}
 \includegraphics{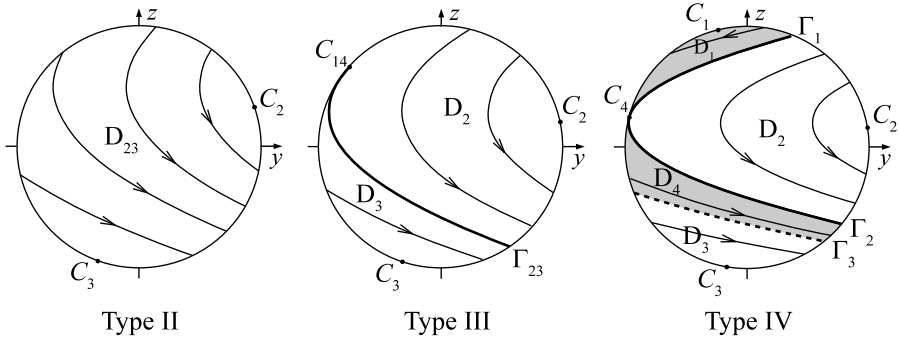}
 \end{center}
 \caption{Three types of the Colombo top established by the values of $a$ and $b$, shown at the $x \geq 0$ hemisphere of S1.
 The discriminant $\Delta$ (\ref{Delta}) is positive in the shaded area, negative in white zones, and null on the thick curves $\Gamma$ or at the Cassini states $C$.
 Exemplary parameters values are: $a=b=0.5$ (type II), $a= (3/4)^3, b = (7/16)^\frac{3}{2}$ (type III), and $a=b=0.2$ (type IV).}
 \label{fig:2}
\end{figure}

\subsubsection{Type II}

When $a^{\frac{2}{3}}+b^{\frac{2}{3}} > 1$, the dynamics on a sphere is relatively simple. There are two stable equilibria (elliptic points):
$C_3$ at the lower ($z<0$) hemisphere and $C_2$ at the upper hemisphere, located at
\begin{equation}\label{z3:Cas}
     z^\ast_3 = \frac{b}{2} - \sqrt{ A_2 } - \sqrt{ B_2},  \qquad  z^\ast_2 = \frac{b}{2} - \sqrt{A_2} + \sqrt{B_2},
\end{equation}
where
\begin{eqnarray}
  A_2 &=&  - P'_2 +\frac{1}{2}\,\left(  \sqrt[3]{g'_3+ \sqrt{-\Delta'} } + \sqrt[3]{g'_3 - \sqrt{-\Delta'} } \right), \nonumber\\
  B_2 & =& -3 P'_2 - A_2 + 2 \sqrt{A_2^2 + 3 P'_2 \left(P'_2+A_2\right) -  \frac{3 \rho^2}{16}},
\end{eqnarray}
and, according to Eqs.~(\ref{deltap}) and (\ref{invp})
\begin{equation}\label{pomcII}
    \sqrt{-\Delta'} = \frac{ab}{4}\sqrt{a^2b^2 -  \rho^3},\quad g'_3 = \frac{a^2 b^2}{4} - \left(\frac{\rho}{2}\right)^3,
    \quad P'_2 = - \frac{\rho}{2} - \frac{b^2}{4}.
\end{equation}
We can identify the Cassini states coordinates as $z^\ast_3 = z_{12}(E_3)$, and $z^\ast_2 = z_{12}(E_2)$, where $z_{12}$ is given by Eq.~(\ref{zm12:f}).

The energy values in the type II problem are bounded by  $E_3 \leq E \leq E_2$.
All trajectories with energy $E_3 < E < E_2$ are simple periodic curves
(one for each value of $E$), oscillating in $z$ between $z_1$ and $z_2$, as given by Eqs.~(\ref{z122:f}).
Extending and modifying the domains labeling of \citet{HenMur:1987}, let us label the entire sphere surface (with the two Cassini states excluded)
as $\mathrm{D}_{23}$ (Fig.~\ref{fig:2}, left).

\subsubsection{Type IV}

When $a^{\frac{2}{3}}+b^{\frac{2}{3}}<1$, the flow is shaped by the presence of three elliptic fixed points $C_1$, $C_2$, $C_3$, and a hyperbolic
point $C_4$ (Fig.~\ref{fig:2}, right). The latter is accompanied by two homoclinic orbits -- upper $\Gamma_{1}$ and lower $\Gamma_{2}$. Thus the sphere
is first partitioned into three domains:
$\mathrm{D}_1$ between $C_1$ and $\Gamma_{1}$, $\mathrm{D}_2$ between $C_2$, $\Gamma_{1}$, $\Gamma_{2}$, and the third area bounded by $C_3$ and $\Gamma_2$.
But the last of the three domains is further divided by the curve whose energy equals that of $C_1$ -- the thick dashed curve $\Gamma_3$ in Fig.~\ref{fig:2}. The
subdomains $\mathrm{D}_3$ and $\mathrm{D}_4$ may look similar from the geometrical point of view, but note the fact that each curve in $\mathrm{D}_4$
has a companion in $\mathrm{D}_1$ with the same energy, which is not the case in $\mathrm{D}_3$.

Equations (\ref{z1234cas}) from Appendix~\ref{Ap:Cas} can be used as they are, but here we add an alternative form, based upon the transformation (\ref{X23h}).
\begin{eqnarray}
  z^\ast_1  &=& \frac{b}{2} + \sqrt{A_4} + \sqrt{B_4} - \sqrt{B'_4} = \frac{b}{2} + \sqrt{A_4} + \sqrt{B_4+B'_4 - 2 \sqrt{B_4 B'_4}}, \nonumber \\
  z^\ast_2  &=& \frac{b}{2} - \sqrt{A_4} + \sqrt{B_4} + \sqrt{B'_4} = \frac{b}{2} - \sqrt{A_4} + \sqrt{B_4+B'_4 + 2 \sqrt{B_4 B'_4}}, \nonumber \\
  z^\ast_3  &=& \frac{b}{2} - \sqrt{A_4} -\sqrt{B_4} - \sqrt{B'_4} = \frac{b}{2} - \sqrt{A_4} - \sqrt{B_4+B'_4 + 2 \sqrt{B_4 B'_4}},  \label{Cas1234} \\
  z^\ast_4  &=& \frac{b}{2} + \sqrt{A_4} - \sqrt{B_4} + \sqrt{B'_4} = \frac{b}{2} + \sqrt{A_4} - \sqrt{B_4+B'_4 - 2 \sqrt{B_4 B'_4}}, \nonumber
\end{eqnarray}
where
\begin{equation}
 A_4 = \frac{b^2}{4}+ \rho \cos^2\frac{\phi_4}{3}, \qquad B_4 =  \frac{b^2}{4}+ \rho \cos^2\frac{\phi_4-\pi}{3} ,
 \qquad B'_4 =  \frac{b^2}{4}+ \rho \cos^2\frac{\phi_4+\pi}{3},
\end{equation}
and
\begin{equation}
\cos{\phi_4} = a b \rho^{-\frac{3}{2}}.
\end{equation}
From $A_4 < B_4 < B'_4$ we infer $z^\ast_3 < 0 < z^\ast_2 < b <  z^\ast_4  < z^\ast_1$. The associated $y$ coordinates satisfy
inequalities $y^\ast_4 < y^\ast_1 < -a <  y^\ast_3 < 0  < y^\ast_2$.

The energy values are $E_3 < E_1 < E_4 < 0 < E_2$. Referring to Sect.~\ref{inters}, we identify
$z^\ast_1 = z_{34}(E_1)$, $z^\ast_2 = z_{12}(E_2)$, $z^\ast_3 = z_{12}(E_3)$, and $z^\ast_4 = z_{23}(E_4)$.
The unstable equilibrium energy $E_4$ is of special importance, because it serves to determine
the turning points of the separatrices from Eq.~(\ref{z14q:f}): $z_4(E_4)$ for $\Gamma_1$, and
$z_1(E_4)$ for $\Gamma_2$.

For the energy values $E_1 < E < E_4$, when $\Delta > 0$, each $E$ refers to two periodic curves: one in $\mathrm{D}_4$ with the turning points $(z_1,z_2)$ given by (\ref{z12:f}), and one in
$\mathrm{D}_1$ with $(z_3,z_4)$ given by Eq.~(\ref{z34:f}). An energy in $E_3 <   E < E_1$ defines only one periodic trajectory in $\mathrm{D}_3$,
and each $E_4 < E < E_2$ defines one periodic curve in $\mathrm{D}_2$. Since $\Delta < 0$ in both cases, the turning points are given by
$(z_1,z_2)$ from Eq.~(\ref{z122:f}).

\subsubsection{Type III}

The case of $a^\frac{2}{3}+b^\frac{2}{3}=1$ is specific, but it has to be included to understand the bifurcation between the two neighboring types.
Increasing $a$ and/or $b$ from the type IV, we observe that the two Cassini states $C_1$ and $C_4$ merge
into a single point $C_{14}$ of neutral stability, the homoclinic orbit $\Gamma_1$ contracts to a point, and the separatrix $\Gamma_2$ merges with
the specific periodic orbit $\Gamma_{3}$ into $\Gamma_{23}$. Thus the domains $\mathrm{D}_1$ and $\mathrm{D}_4$ disappear. In course of
transition from type III to type II, the division between $\mathrm{D}_{2}$ and $\mathrm{D}_3$ disappears, hence we merge them into a single $\mathrm{D}_{23}$.
\citet{LanEli:1995} refer to this transition under the name of teardrop bifurcation.

The coordinates of the Cassini states can be obtained from Eq.~(\ref{Casdeg}) and are expressible in terms of $a$ or $b$ alone, resulting in
\begin{equation}\label{yz14}
     z^\ast_{14} = \sqrt{1-A_3}= b^\frac{1}{3}, \qquad y^\ast_{14} = - \sqrt{A_3} = - a^\frac{1}{3},
\end{equation}
and
\begin{equation}\label{zt3:Cas}
    z^\ast_3 = - z^\ast_{14} \left(A_3 + \sqrt{B_3} \right), \qquad  z^\ast_2 = - z^\ast_{14} \left(A_3 - \sqrt{B_3} \right),
\end{equation}
where
\begin{equation}\label{AB3}
    A_3 = 1- b^\frac{2}{3} = a^\frac{2}{3}, \qquad B_3 = A^2_3-A_3+1.
\end{equation}

The expression of the energy at $C_{14}$ is so simple, that we write it explicitly
\begin{equation}\label{E14}
 E_{14} = - \frac{3}{2} \sqrt[3]{a^4 b^2}= - \frac{3 \left( z^\ast_{14}\right)^2  \left(y^\ast_{14}\right)^4}{2} = - \frac{3 A_3^2 \left(1-A_3\right)}{2}.
\end{equation}
It is useful also in finding the turning point of $\Gamma_{23}$ at $x=0$, which is
\begin{equation}\label{zmym3}
    z_\mathrm{m} = z^\ast_{14} \left(1-4A_3\right), \qquad  y_\mathrm{m} = - y^\ast_{14} \left(3-4 A_3\right).
\end{equation}

For regular trajectories with  $E_3 < E  < E_2$, and $E \neq E_{14}$,  their turning points are given by $z_1$ and $z_2$ from Eq.~(\ref{z122:f}),
which also means $z_1 < z_2$.

\section{Analytical solution}
\label{ansol}
\subsection{Weierstrass form}

\label{Weier}
\subsubsection{General framework}
\label{Wgf}

Equations of motion (\ref{eqm:c}) or (\ref{eqm:C}) admit an exact analytical  solution in terms of the elliptic functions.
Actually, the solution hinges upon the fact that the variable $Z$ can be considered separately from the remaining two.
To see it, we can differentiate $\dot{Z}=-aX$,
substitute $\dot{X}$ from the first of Eqs.~(\ref{eqm:C}), and use the energy integral (\ref{Hxyzs}) to eliminate $Y$, obtaining
\begin{equation}\label{Z2d}
    \ddot{Z} = - \frac{1}{2} Z^3 - E Z - a^2 b.
\end{equation}
Equation (\ref{Z2d}) defines a 1 degree of freedom, conservative system with the potential
\begin{equation}\label{Vzdef}
    V_z(Z) = \frac{Z^4}{8} + \frac{E}{2} Z^2 + a^2 b Z,
\end{equation}
and the energy integral
\begin{equation}\label{EnZ}
    \frac{\dot{Z}^2}{2} + V_z(Z) = E_z = \mathrm{const}.
\end{equation}

But the potential $V_z(Z)$ and the energy $E_z$ are closely related to the polynomial $W(Z)$:
\begin{equation}
E_z - V_z(Z) = \frac{1}{8} W(Z), \mbox{~~with~~} 2 E_z = \frac{W(0)}{4} = -  \left(E - a^2\right)^2 +  a^2 \left(1- b^2\right).
\end{equation}
Thus, whether we use Eq.~(\ref{EnZ}), or the squared Eqs.~(\ref{eqm:C}) and (\ref{XofZ}),
the outcome is $4 \dot{Z}^2 = W(Z)$, amenable to the separation of variables method.

Since $W(Z)$ is a quartic polynomial, finding $Z(t)$ amounts to inverting the elliptic integral
\begin{equation}\label{integr}
  \sigma_z  \intop_{Z_{0}}^{Z}\frac{\rd Z}{\sqrt{W(Z)}} =  \frac{1}{2} \intop_{t_0}^{t}  \rd t,
\end{equation}
where $\sigma_z = \sgn{\dot{Z}} = - \sgn{X}$, and the right-hand side evaluates to
\begin{equation}\label{taudef}
    \tau_0 = \frac{t-t_0}{2}.
\end{equation}
The integral to the left of (\ref{integr}) should be a monotonous function of $Z$ to allow the inversion
 (solving for $Z(t)$). In the admissible range of $Z$ between two turning points (or one turning point and an unstable equilibrium)
 the sign of $\dot{Z}$ is constant, which allows to establish its value from the initial condition and to pull $\sigma_z$ out of the integrand.

 The integral to the left of Eq.~(\ref{integr}) is an elliptic integral and it can be reduced to the Weierstrass normal form
 by an appropriate transformation $Z \rightarrow s$, so that
\begin{equation}\label{int:Wan}
   \sigma_z \intop_{Z_0}^{Z}\frac{\rd Z}{\sqrt{W(Z)}} =     \intop_{s}^{\infty} \frac{\rd s}{\sqrt{S(s)}} = \tau_0,
\end{equation}
where $S(s)= 4 s^3 - g_2 s - g_3,$ is the cubic polynomial of the Weierstrass resolvent (\ref{Sdef}) for $W(Z)=0$,
with the invariants $g_2$, $g_3$ defined in Appendix~\ref{Ap:Wr} -- Eqs.~(\ref{g2}) and (\ref{g3}). Note that the
initial value $Z_0$ is always mapped to $s \rightarrow \infty$,
regardless of the ordering of the integration limits $Z_0$ and $Z$, which explains the presence of $\sigma_z$ in the forthcoming transformations.

Solving the rightmost part of Eq.~(\ref{int:Wan}) amounts to the substitution of the Weierstrass elliptic function
\begin{equation}
 s = \wp(\tau_0; g_2, g_3), \mbox{~~~~hence~~~~}  \rd s = \wp'(\tau_0; g_2, g_3) \rd \tau_0.
\end{equation}
In all further instances we will use the abbreviated notation for the Weierstrass $\wp$ function
$\wp(u) = \wp(u; g_2, g_3)$, whenever the invariants $g_2$ and $g_3$ from Appendix~\ref{Ap:Wr} are used.
Only the invariants different than $g_2$ and $g_3$ will be added to the list of arguments if needed.
The derivative od the Weierstrass $\wp$ function obeys
\begin{equation}\label{derWe}
    \left| \wp'(\tau_0 ) \right| = \sqrt{S(\wp(\tau_0 ))},
\end{equation}
and in the first half-period $0 < \tau_0 < \omega_1 =  \wp^{-1}(e_1)$, we have $\wp'(\tau_0 )<0$. Thus, indeed
\begin{equation}\label{solds}
     \intop_{s}^{\infty} \frac{\rd s}{\sqrt{S(s)}} =  \intop_{\tau_0}^{0} \frac{\wp'(\tau)}{\left| \wp'(\tau)\right|}\,\rd \tau
     = - \intop_{\tau_0}^{0} \rd \tau = \tau_0,
\end{equation}
as expected from Eq.~(\ref{int:Wan}).

Once the solution for the first half-period is found, its continuation can be investigated by the substitution into
Eq.~(\ref{Z2d}) which is free from the restrictions imposed by the inversion procedure.

\subsubsection{Initial conditions at turning point}

Let us begin with the easiest situation, when the initial condition is $Z_0 = Z_j = z_j-b$, and $t_0=t_j$ is the epoch of crossing
the turning point $X_0=0$, i.e. $\dot{Z}=0$. Then the integral to the left of (\ref{int:Wan}) is reduced to the Weierstrass normal form
in variable $s$ through the rational transformation \citep{Enneper:1890,Bianchi:1901}
\begin{equation}\label{Bia:Zs}
    Z = Z_j + \frac{\frac{1}{4} W'(Z_j)}{s - \frac{1}{24} W''(Z_j)}, \qquad s = \frac{W'(Z_j)}{4 \left(Z-Z_j\right)} +  \frac{1}{24} W''(Z_j).
\end{equation}
Replacing the subscript $0$ with $j$ in the formulae of Sect.~\ref{Wgf}, we find
\begin{equation}\label{Bia:ZW}
    Z = Z_j + \frac{\frac{1}{4} W'(Z_j)}{\wp(\tau_j) - \frac{1}{24} W''(Z_j)},
\end{equation}
where
\begin{equation}\label{taujdef}
    \tau_j = \frac{t-t_j}{2}.
\end{equation}
Thanks to the symmetry of trajectory with respect to the turning point, the solution does not depend on $\sigma_z$, and remains valid for all values of $\tau_j$.

Given the initial conditions
$X(t_j) = 0$, $Y(t_j)=Y_j$, $Z(t_j)=Z_j$, and knowing $Z(t)$, we may complete the solution for $\vv{R}(t)$. The two missing variables come from the energy integral (\ref{Hxyzs}), with
\begin{equation}\label{Ener0t}
    E = - \frac{Z^2}{2} + a Y = - \frac{Z_j^2}{2} + a Y_j,
\end{equation}
and from the third of the equations of motion (\ref{eqm:C}), i.e. $X = - \dot{Z}/a$, so
\begin{eqnarray}
    X & = & - \left(\frac{Z -Z_j}{2 a}\right)^2 \frac{\wp'(\tau_j)}{a b + Y_j Z_j}, \nonumber \\
    Y & = &  Y_j + \frac{Z^2-Z_j^2}{2 a}, \label{soltj} \\
    Z & = &  Z_j - \frac{ 2 a^2 b + 2 E Z_j + Z_j^3}{\wp(\tau_j) + \frac{1}{3} E + \frac{1}{2} Z_j^2} =
   Z_j - \frac{6 a \left(a b + Y_j Z_j\right)}{3 \wp(\tau_j) + a Y_j + Z_j^2}. \nonumber
\end{eqnarray}
The coordinates on the unit sphere are, as usually, $x=X$, $y=Y-a$, and
$z=Z+b$.

\subsubsection{Arbitrary initial conditions: Weierstrass--Biermann form}

\label{WBF}
If the initial conditions are not at the turning point, i.e. $Z_0 \neq Z_j$,
the transformation $Z \mapsto s$ is more cumbersome than (\ref{Bia:Zs}). \citet{WhWa:1927}
quote two alternative forms of the final solution: one due to Weierstrass, published by \citet{Biermann:1865}, and one by \citet{Mordell:1915}.
Actually, there is yet another, formally elegant form -- the \textit{secondo metode d'inversione} of \citet{Bianchi:1901}, but the relation of
its constants to initial conditions is rather awkward, so we do not consider it here.

The Weierstrass-Biermann form results from the substitution \citep{Enneper:1890}
\begin{eqnarray}\label{W-B}
    Z & = & Z_0 + \frac{\sigma_z \sqrt{W_0} \sqrt{S(s)} +\frac{1}{2} W_1 \left(s - \frac{1}{12} W_2\right) +
    \frac{1}{4} W_0 W_3}{2 \left(s-\frac{1}{12} W_2\right)^2 - \frac{1}{2} W_0 W_4 }, \\
    \label{W-Bi}
    s & = & \frac{\sqrt{W_0} \sqrt{W(Z)} + W_0 + \frac{1}{2} W_1 \left(Z-Z_0\right)
    + \frac{1}{6} W_2 \left(Z-Z_0\right)^2 }{2 \left( Z-Z_0 \right)^2},
\end{eqnarray}
where $\sigma_z$ is the sign of  $Z-Z_0$, and
\begin{equation}
     W_k = \frac{1}{k!} \left[\frac{\rd^k W(Z)}{\rd Z^k}\right]_{Z_0},
\end{equation}
are the Taylor coefficients in
\begin{equation}
W(Z) = \sum_{k=0}^4 W_k (Z-Z_0)^k.
\end{equation}
After expressing $E$ in terms of the initial conditions, they take the form
\begin{eqnarray}
  & &   W_0 = 4 a^2 X_0^2, \qquad \qquad \, \, W_1 = - 8 a \left(a b+Y_0 Z_0\right), \nonumber \\
  & &  W_2 = - 4 \left(a Y_0 + Z_0^2\right), \quad W_3 = - 4 Z_0, \quad W_4 = -1. \label{Wkdef}
\end{eqnarray}

Substituting $s = \wp(\tau_0)$, and recalling that $\sqrt{S(s)} = - \wp'(\tau_0)$ over the first half-period,  we obtain
\begin{equation}\label{W-B:1}
    Z = Z_0 + \frac{-\sigma_z \sqrt{W_0} \wp'(\tau_0) +\frac{1}{2} W_1 \left(\wp(\tau_0) - \frac{1}{12} W_2\right) + \frac{1}{4} W_0 W_3}{
    2 \left(\wp(\tau_0) -\frac{1}{12} W_2\right)^2 - \frac{1}{2} W_0 W_4 }.
\end{equation}
Then, accounting for Eqs.~(\ref{Wkdef}) and the fact that $\sqrt{W_0} = 2 a |X_0| = - 2 \sigma_z  a X_0$,
\begin{equation}\label{W-B:Z}
    Z = Z_0 + a \, \frac{X_0 \wp'(\tau_0) - 2 \left(a b + Y_0 Z_0 \right) \left[\wp(\tau_0) + \frac{1}{3} \left(a Y_0 + Z_0^2\right)\right]
    - 2 a X_0^2 Z_0}{
    \left[\wp(\tau_0) + \frac{1}{3} \left(a Y_0 + Z_0^2\right)\right]^2 + a^2 X_0^2 }.
\end{equation}
The substitution into Eq.~(\ref{Z2d}) with the initial condition $\dot{Z}_0 = - a X_0$, shows that the formula (\ref{W-B:Z}) is actually
valid for all values of $\tau_0$, so the initial restriction to the first half-period can be abolished.

Like before, the remaining two variables $Y,X$ are found from the energy integral  and from the equations of motion, respectively.
Thus for $Y$ we have simply
\begin{equation}\label{W-B:Y}
    Y = Y_0 + \frac{Z^2-Z_0^2}{2 a},
\end{equation}
whereas $X= - \dot{Z}/a$, requires the differentiation and some manipulations leading to
\begin{equation}\label{W-B:X}
    X =  \frac{- X_0 \left( 3 \wp(\tau_0)- \frac{1}{4}g_2\right) + \left(a b + Y_0 Z_0 \right) \wp'(\tau_0)
    + \frac{(Z-Z_0)}{a} \left[\wp(\tau_0) + \frac{1}{3} \left(a Y_0 + Z_0^2\right)\right] \wp'(\tau_0)}{
    \left[\wp(\tau_0) + \frac{1}{3} \left(a Y_0 + Z_0^2\right)\right]^2 + a^2 X_0^2 },
\end{equation}
where the second derivative has been removed using the identity $\wp''=6 \wp^2-(g_2/2)$.

\subsubsection{Arbitrary initial conditions: Safford form}
\label{Safsec}

If $X_0=0$, so $\tau_0=\tau_j$, the above solution simplifies to Eqs.~(\ref{soltj}) in a straightforward manner.
On the other hand, as pointed out by \citet{Safford:1919}, the general solution (\ref{W-B:1}) can be derived from the particular solution (\ref{Bia:ZW})
by assuming $\tau_j = \tau_0 + \phi_j$, and making use of the addition theorem for the Weierstrass function $\wp$.
Given the initial conditions $X_0,Y_0,Z_0$ at $t=t_0$, we can find the turning point coordinates $Z_j, Y_j$ for this trajectory using the formulae
of Sect.~\ref{inters}. Then, according to \citet{Safford:1919}, the phase $\phi_j$ is defined through Eq.~(\ref{W-Bi}) at $Z=Z_j$, giving
$s=s_j = \wp(\phi_j)$.
Thus, as an alternative to using Eqs.~(\ref{W-B:Z}) and (\ref{W-B:Y}) for arbitrary initial conditions, one can first compute $\phi_j = \wp^{-1}(s_j)$,
and then apply the particular solution (\ref{soltj}) with $\tau_j = \tau_0 + \phi_j$.

But \citet{Safford:1919} cared solely about the reduction of the integrand form, so he paid no attention to the problem that
$\wp(\phi_j)$ is uniquely invertible only in the domain $0 \leq \phi_j \leq \omega_1$, i.e. within the first half-period.
In order to properly place the phase in the full period range  $- \omega_1 \leq \phi_j \leq \omega_1$, one needs the information about the sign of
the derivative $\wp'(\phi_j)$.
To this end, we take a slightly different approach, that actually leads to a simpler expression for $s_j$.
Substituting $\tau_j=\phi_j$, and $\vv{R}=\vv{R}_0$ in Eqs.~(\ref{soltj}), we can solve them to find
\begin{eqnarray}
  \wp(\phi_j) &=& - 2 a \frac{a b + Y_j Z_j}{Z_0-Z_j} - \frac{a Y_j + Z_j^2}{3}, \label{sj:1} \\
  \wp'(\phi_j) &=& -  X_0 \left(a b + Y_j Z_j\right) \, \left[\frac{2a}{Z_0-Z_j}\right]^2, \label{spj:1}
\end{eqnarray}
and this set allows the unique determination of the phase. Then, we can apply Eqs.~(\ref{soltj}) to obtain
\begin{equation} \label{solSaf}
    X= - \left(\frac{Z -Z_j}{2 a}\right)^2 \frac{\wp'(\tau_0+\phi_j)}{a b + Y_j Z_j}, \qquad
    Z =
   Z_j - \frac{6 a \left(a b + Y_j Z_j\right)}{3 \wp(\tau_0+\phi_j) + a Y_j + Z_j^2},
\end{equation}
with $Y_j$ and $Y$ derived from the energy integral.

Compared to the Weierstrass-Biermann solution, we gain the simplicity at the expense of pre-computing the turning point coordinates
and the phase for the given (arbitrary) initial conditions.

\subsubsection{Arbitrary initial conditions: Mordell form}

The inversion formula of \citet{Mordell:1915} was formulated in the language of
homogeneous binary forms; in order to apply it to the Colombo top problem, let us translate it to the univariate polynomials framework.
The link is simple: the quartic polynomial $W(Z)$ from Eq.~(\ref{Wagen}),
and the quartic binary form
\begin{equation}\label{Vagen}
    V(\xi,\eta) = a_0 \xi^4 + 4 a_1 \xi^3 \eta + 6 a_2 \xi^2 \eta^2 + 4 a_3 \xi \eta^3 + a_4 \eta^4,
\end{equation}
can be matched by
\begin{equation}\label{WVlink}
    W(Z) = V(Z,1), \qquad V(\xi,\eta) = \eta^4 W(\xi \eta^{-1}).
\end{equation}
Thus, after the substitution $Z=\xi \eta^{-1}$, Eq.~(\ref{int:Wan}) is equivalent to
\begin{equation}\label{integr:M}
     \sigma_z \intop_{(\xi_0,\eta_0)}^{(\xi,\eta)}\frac{\eta \rd \xi - \xi \rd \eta}{\sqrt{V(\xi,\eta)}}
    =     \intop_{s}^{\infty} \frac{\rd s}{\sqrt{S(s)}} = \tau_0,
\end{equation}
where the definite integral in $Z$ has been replaced by a path-independent line integral.
Skippping the intermediate steps described in \citep{Mordell:1914, Mordell:1915}, the inversion of (\ref{integr:M}) results in
\begin{eqnarray}
  \xi &=& -\sigma_z \xi_0 \sqrt{V_0} \wp'(\tau_0) + \frac{1}{2} \frac{\partial V_0}{\partial \eta_0} \wp(\tau_0)
  + \frac{1}{2} \frac{\partial \tilde{h}_0}{\partial \eta_0}, \nonumber  \\
  \eta &=& -\sigma_z \eta_0 \sqrt{V_0} \wp'(\tau_0) - \frac{1}{2} \frac{\partial V_0}{\partial \xi_0} \wp(\tau_0)
  - \frac{1}{2} \frac{\partial \tilde{h}_0}{\partial \xi_0}, \label{xiet}
\end{eqnarray}
where $V_0 = V(\xi_0,\eta_0)$, and $\tilde{h}_0$ is the Hessian covariant\footnote{In this paper we use the Hessian covariant as defined by \citet{Jans:2011}, which is the same as
in \citet{WhWa:1927}. Its sign is opposite to the one originally applied by \citet{Mordell:1915}.} of $V$.

From the correspondence rules (\ref{WVlink}) we derive
\begin{eqnarray}
\frac{\partial V_0}{\partial \xi_0}   &=&  \eta_0^3 W_1, \nonumber \\
\frac{\partial V_0}{\partial \eta_0} &=& \eta_0^2 \left(4 \eta_0 W_0 - \xi_0 W_1\right), \nonumber \\
\tilde{h}_0 & = & \frac{1}{144} \left(\frac{\partial^2 V_0}{\partial \xi_0^2} \frac{\partial^2 V_0}{\partial \eta_0^2}
- \left(\frac{\partial^2 V_0}{\partial \xi_0 \partial \eta_0} \right)^2\right) = \frac{\eta_0^4 \left(8 W_0 W_2 - 3 W_1^2\right)}{48}, \label{Vhdef}\\
\frac{\partial \tilde{h}_0}{\partial \xi_0}   &=& \frac{\eta_0^3 \left(6 W_0 W_3 - W_1 W_2\right)}{12}, \nonumber \\
\frac{\partial \tilde{h}_0}{\partial \eta_0}   &=& \frac{\eta_0^2 \left(8 \eta_0 W_0 W_2 - 3 \eta_0 W_1^2
+ \xi_0 W_1 W_2 - 6 \xi_0 W_0 W_3\right)}{12}, \nonumber
\end{eqnarray}
where $W_n$ are defined in Eq.~(\ref{Wkdef}). By letting $Z=\xi/\eta$, and $Z_0=\xi_0/\eta_0$, we find that
(\ref{xiet}) and (\ref{Vhdef}) amount to
\begin{equation}\label{Mor:f}
    Z = \frac{\xi}{\eta} = Z_0 + \frac{48 W_0 \wp(\tau_0) + 8 W_0 W_2 - 3 W_1^2}{
    - 24 \sigma_z \sqrt{W_0} \wp'(\tau_0) - 12 W_1 \wp(\tau_0)+ W_1 W_2 - 6 W_0 W_3}.
\end{equation}
Proceeding like in Sect.~\ref{WBF} we obtain the final form
\begin{equation}\label{Mor:ff}
    Z =  Z_0 +  4 a \,  \frac{ X_0^2 \left[  \wp(\tau_0)  - \frac{2}{3} \left(a   Y_0 + Z_0^2\right)\right] -   \left( a b + Y_0 Z_0\right)^2 }{
  X_0 \wp'(\tau_0) + 2 \left(a b +  Y_0 Z_0\right) \left[  \wp(\tau_0) + \frac{1}{3}\left( a Y_0 + Z_0^2\right)\right] + 2 a X_0^2 Z_0},
\end{equation}
looking different from the Weierstrass-Biermnann solution (\ref{W-B:Z}), yet providing the same values of $Z$. Notably, the Mordell
solution involves only the first power of $\wp$.

In order to demonstrate the equivalence of the Weierstrass-Biermann and the Mordell solutions, one can multiply the numerator an the denominator
in Eq.~(\ref{W-B:Z}) by the factor
\[
    \left( 48 a X_0 \wp'(\tau_0) - \left( 12 \wp(\tau_0) - W_2 \right) W_1 - 6 W_0 W_3 \right),
\]
use the identity $\left( \wp'(\tau_0) \right)^2 = S(\wp(\tau_0))$,
and substitute the expressions of the invariants in terms of $W_n$
\begin{equation}\label{g23W}
    g_2 = \frac{W_2^2}{12} - \frac{W_2 W_3}{4} + W_0 W_4, \qquad g_3 =   \left|
            \begin{array}{ccc}
              W_4 & \frac{1}{4} W_3 & \frac{1}{6} W_2 \\
              \frac{1}{4} W_3 & \frac{1}{6} W_2 & \frac{1}{4} W_1 \\
              \frac{1}{6} W_2 & \frac{1}{4} W_1 & W_0 \\
            \end{array}
          \right|,
\end{equation}
by analogy with the definitions (\ref{g2}) and (\ref{g3}). The result of this procedure is the Mordell solution (\ref{Mor:ff}).
Obviously, both (\ref{W-B:Z}) and (\ref{Mor:ff}) admit the same limit expression (\ref{soltj}) when $X_0=0$ and $\tau_0=\tau_j$.

As usually, the solution for $X$ and $Y$ can be derived from the equations of motion and the energy integral, like in Sect.~\ref{WBF}.

\subsection{Solution in terms of the Jacobi elliptic functions}

\label{Jacobi:sec}

\subsubsection{Weierstrass functions in terms of the Jacobi functions }

Although the solution in terms of the Weierstrass $\wp$ function presented in Sect.~\ref{Weier} may look universal, its qualitative
properties depend on the values of the invariants through the sign of the discriminant $\Delta$. Indeed, the sign plays the central role in
expressing the solution in terms of the Jacobian elliptic functions. In this section, only the generic, $\Delta \neq 0$ cases are to be discussed.

The basic relation between the Weierstrass and Jacobi functions is formally universal \citep{BF:1971}
\begin{eqnarray}\label{WJa}
    \wp(\tau) & = & e_3 + \left(\frac{\gamma_p}{\sn(u_p,k_p)}\right)^2 = e_3 + \gamma_p^2 \, \frac{1+\dn(2 u_p,k_p)}{1- \cn(2 u_p,k_p)} , \\
     \wp'(\tau) & = & - 2 \left(\frac{\gamma_p}{\sn(u_p,k_p)}\right)^3 \cn(u_p  ,k_p)\,\dn(u_p ,k_p)  \nonumber\\
    & = & - 2 \gamma_p^3 \, \frac{\left(\cn(2 u_p,k_p)+\dn(2 u_p,k_p)\right) \left(1+\dn(2 u_p,k_p)\right)}{\sn(2 u_p,k_p) \left(1- \cn(2 u_p,k_p)\right)},
\label{WpJa}
\end{eqnarray}
where
\begin{equation}\label{ukg}
  u_p = \gamma_p \tau, \qquad  \gamma_p = \sqrt{e_1-e_3}, \qquad   k_p = \sqrt{\frac{e_2-e_3}{e_1-e_3}}.
\end{equation}
But if we restrict considerations to the real arguments and moduli of the elliptic functions, the above
expressions are valid only if the discriminant $\Delta$ from Eq.~(\ref{Delta}) is positive.

When $\Delta < 0$, which means complex $e_2$ and $e_3$, the appropriate form is  \citep{AbraSteg}
\begin{eqnarray}\label{WJam}
    \wp(\tau) & = &  =  e_1 +  \gamma_n \,\left(\frac{\cn(u_n,k_n)}{ \sn(u_n, k_n) \dn(u_n, k_n)}\right)^{2} =
    e_1 +  \gamma_n \, \frac{1+\cn(2 u_n, k_n)}{1-\cn(2 u_n, k_n)}, \\
     \wp'(\tau) & = & -2 \left(\frac{ \sqrt{\gamma_n} }{\sn(u_n, k_n) \dn(u_n, k_n)} \right)^3 \, \cn(u_n,k_n) \, \left( 1-k_n^2 + k_n^2 \cn^4(u_n,k_n) \right)
     \nonumber \\
     &=& - 4 \gamma_n^\frac{3}{2} \, \frac{\sn(2 u_n, k_n) \dn(2 u_n, k_n)}{\left(1 - \cn(2 u_n,k_n)\right)^2}, \label{WpJma}
\end{eqnarray}
where
\begin{equation}
    u_n = \sqrt{\gamma_n} \tau, \qquad  \gamma_n = \gamma_p \sqrt{e_1-e_2}   = \frac{1}{2} \sqrt{12 e_1^2 - g_2}, \qquad
     k_n = \frac{1}{2} \sqrt{2 - \frac{3 e_1}{\gamma_n} }. \label{unkn}
\end{equation}
The two cases are linked by the complex modulus transformation -- see \citet[][formula 165.07]{BF:1971}.

Let the initial conditions at the epoch $t_0$ be $(X_0,Y_0,Z_0)$.
The equations relating the Jacobi and Weierstrass functions can be substituted in to any of the solution forms provided in Sect.~\ref{Weier}.
For the Weierstrass--Biermann or the Mordell form, it is enough to compute the energy $E=E(Y_0,Z_0)$ and the discriminant
$\Delta$ to choose the appropriate set (\ref{WJa},\ref{WpJa}) or (\ref{WJam},\ref{WpJma}). Then,
from the invariants $g_2, g_3$ the roots $e_1, e_2, e_3$ are found, which allows the computation of $\vv{R}(t)$ or $\vv{r}(t)$ for any epoch $t$.

Below, we discuss the Safford form from Sect.~\ref{Safsec}, which allows to use the initial conditions at any $t_0$, but requires
the turning point coordinates $(X_j=0, Y_j=y_j+a, Z_j = z_j-b)$ as supplementary parameters. Having computed two appropriate turning
points, such that either $Z_1 < Z_0 < Z_2$, or $Z_3 < Z_0 < Z_4$, we pick one of them as the reference point $Z_j$ to be used in Eq.~(\ref{solSaf}).
Then, after determining  phase $\phi_j$ with respect to the turning point $Z_j$, the motion can be computed for any epoch $t$,
using $\tau_j = \tau_0+\phi_j$.

\subsubsection{Motion in $\mathrm{D}_1$ and $\mathrm{D}_4$}
\label{secD1D4}

The case $\Delta > 0$ occurs only in type IV, when the energy is bounded by $E_4 < E < E_1$: either in the domain $\mathrm{D}_1$,
where turning points $Z_3, Z_4$ are given by Eq.~(\ref{z34:f}) with  $Z_4^\ast < Z_3 < Z_4$, or in the domain $D_4$, where the turning points are
 $Z_1, Z_2$ given by Eq.~(\ref{z12:f}) and  $Z_1 < Z_2 < Z_4^\ast$. Then, selecting any appropriate $Z_j$ as the reference point, we can use
 the expressions
 \begin{eqnarray}
   X & = & \frac{A_p \gamma_p \sn(u_p,k_p) \cn(u_p,k_p) \dn(u_p,k_p)}{\left(1+ B_p \sn^2(u_p,k_p)\right)^2}\\
   Y & = & Y_j + \frac{Z^2-Z_j^2}{2 a } = Y_j - \frac{A_p \sn^2(u_p,k_p)}{1+ B_p \sn^2(u_p,k_p)}
   \left( Z_j - \frac{a A_p \sn^2(u_p,k_p)}{ 1+ B_p \sn^2(u_p,k_p)}\right)  \label{solD14} \\
   Z &=& Z_j - \frac{a A_p \sn^2(u_p,k_p)}{1+ B_p \sn^2(u_p,k_p)}
 \end{eqnarray}
 where
 \begin{equation} \label{ABp}
  A_p = \frac{2 (a b + Y_j Z_j)}{e_1-e_3}, \quad B_p = \frac{\frac{1}{3} \left(a Y_j + Z_j^2 \right) + e_3}{e_1-e_3},
 \end{equation}
 and the phase $\phi_j$ in
\begin{equation}
 u_p = \gamma_p (\tau_0 + \phi_j),
\end{equation}
can be computed from the incomplete elliptic function of the first kind F:
\begin{eqnarray}
 \gamma_p \phi_j & = & \mathrm{F}(v_j,k_p), \\
 v_j & = & \am(\gamma_p \phi_j, k_p) = \sgn{(A_p X_0)}\,\arcsin \sqrt{\frac{Z_j-Z_0}{a A_p - (Z_j-Z_0) B_p}}. \label{vjp}
\end{eqnarray}
The motion is periodic and the period $P_t$ (with respect to time $t$) is given by the complete elliptic integral of the first kind K
\begin{equation}\label{perip}
    P_t = \frac{4 \mathrm{K}(k_p)}{\gamma_p}.
\end{equation}

\subsubsection{Motion in $\mathrm{D}_2$, $\mathrm{D}_3$ and $\mathrm{D}_{23}$}

The common feature of trajectories in domains  $\mathrm{D}_2$, $\mathrm{D}_3$ and $\mathrm{D}_{23}$ is the negative discriminant $\Delta$.
Thus, given the initial conditions and resulting energy, we pick $Z_1$ or $Z_2$ computed from Eqs.~(\ref{z122:f}) as the reference point $Z_j$
and compute, for any $t$
 \begin{eqnarray}
   X & = & \frac{A_n \sqrt{\gamma_n} (1-B_n) \sn(2 u_n,k_n)  \dn(2 u_n,k_n)}{\left(1- B_n \cn(2u_n,k_n)\right)^2}\\
   Y & = & Y_j + \frac{Z^2-Z_j^2}{2 a } = Y_j - A_n \frac{ 1-\cn(2u_n,k_n) }{1 - B_n \cn(2u_n,k_n)}
   \left( Z_j - \frac{a A_n}{2} \,\frac{ 1-\cn(2u_n,k_n)}{ 1 - B_n \cn(2u_n,k_n)}\right),  \label{solD23}\\
   Z &=& Z_j - a A_n\, \frac{ 1-\cn(2u_n,k_n)}{ 1 - B_n \cn(2u_n,k_n)},
 \end{eqnarray}
 where
 \begin{equation}
  A_n = \frac{2 (a b + Y_j Z_j)}{e_1+ \gamma_n + \frac{1}{3} \left(a Y_j + Z_j^2 \right) },
  \quad B_n = 1- \frac{2 \gamma_n}{e_1+ \gamma_n + \frac{1}{3} \left(a Y_j + Z_j^2 \right) }.
 \end{equation}
 The phase in
 \begin{equation}
 u_n = \sqrt{\gamma_n} ( \tau_0 + \phi_j),
 \end{equation}
 results from
\begin{eqnarray}
 2 \sqrt{\gamma_n} \phi_j & = & \mathrm{F}(v_j,k_n), \\
 v_j & = & \am(2 \sqrt{\gamma_n} \phi_j, k_n) = \sgn{(A_n X_0)}\,\arccos{\left(\frac{a A_n + Z_0-Z_j}{a A_n + (Z_0-Z_j) B_n} \right)}.
\end{eqnarray}
The related period is
\begin{equation}\label{perin}
    P_t = \frac{4 \mathrm{K}(k_n)}{\sqrt{\gamma_n}}.
\end{equation}

The above formulation is universal, i.e. appropriate in types II, III, and IV, provided $\Delta < 0$.

\subsection{Special cases}
\label{Spec:sec}
\subsubsection{Reduction rules}

By special cases we mean the ones where elliptic functions reduce to the elementary ones. They could be studies by taking limits of
the Jacobi functions at $k_n$ or $k_p$ tending to 0 or 1. But it is more direct to observe that each of the special cases, be it the Cassini states,
separatrices, or the special curve $\Gamma_3$, results from the reduction of the Weierstrass function $\wp(u,g_2,g_3)$ to the special case
\begin{equation}\label{WPspec}
    \wp(u;3,1) = 1 + \frac{3}{2 \tan^2{\left(\sqrt{\frac{3}{2}} u\right)}}.
\end{equation}
Since  $\Delta=0$ implies $(g_2/3)^3 = g_3^2$, reduction to the above form is always possible by the homogeneity relations
\citep{AbraSteg}
\begin{equation}\label{scal}
    \wp(  u ; g_2,g_3) =  \lambda^{-2} \wp(\lambda^{-1} u;  \lambda^4 g_2 , \lambda^6 g_3),
\end{equation}
except for $g_2=g_3=0$, when
\begin{equation} \label{red0}
\wp(u;0,0) = u^{-2}.
\end{equation}
For the derivative $\wp'$, respective equations are obtained by straightforward differentiation.

Depending on the sign of $g_3$, two procedures are available. For $g_3 > 0$, the substitution of $\lambda=g_3^{-\frac{1}{6}}$
leads straight to
\begin{equation}\label{red1}
    \wp(  u ; g_2,g_3) =  e_1 \wp(\sqrt{e}_1\, u;  3 , 1),
\end{equation}
where $g_3=e_1^3$, according to Eq.~(\ref{e1e23}).

If $g_3 <0$, two steps are taken. First, letting $\lambda=\mathrm{i}$, we convert
\begin{equation}\label{red2}
    \wp(  u ; g_2,g_3) = - \wp(\mathrm{i}\, u;  g_2,-g_3).
\end{equation}
Then, with $\lambda= (-g_3)^{-\frac{1}{6}}$, we recover
\begin{equation}\label{red3}
    \wp(  u ; g_2,g_3) = - 2 e_{12}  \wp(\mathrm{i}\,\sqrt{2 e_{12}} \, u;  3, 1),
\end{equation}
according to Eq.~(\ref{e12e3}).

\subsubsection{Cassini states $C_2$ and $C_3$}

When discussing the Cassini states, we can use the simplest form (\ref{soltj}) for $Z$.
Although the time dependence at the Cassini states does vanish due to $a b+ Y_0 Z_0 = -W'(Z_0)/(8a) = 0$, and $X_0=0$,
it remains of interest to inspect $\wp(\tau)$ in the numerator, because its period is the period of small oscillations
around the stable equilibrium.

Given $a$ and $b$, one should first establish the problem type in order to compute the appropriate coordinate $Z^\ast_2 = z^\ast_2-b$,
or $Z^\ast_3 = z^\ast_3-b$. These are Eqs.~(\ref{z3:Cas}), (\ref{zt3:Cas}), and (\ref{Cas1234}) for the types II, III, and IV, respectively.
Starting from this point, the procedure is common: $Z^\ast_j$ gives the energy of the Cassini state
\begin{equation}\label{EofZ}
    E_j = - \frac{\left(Z^\ast_j\right)^3 + a^2 b}{2 Z^\ast_j}, \qquad j=2,3,
\end{equation}
which substituted in Eq.~(\ref{g2}) or (\ref{g3}) gives the invariants $g_2(E_j)$ or $g_3(E_j)$ -- both positive.

Resorting to Eqs.~(\ref{red1}) and (\ref{WPspec}), we find that $Z$ in solution (\ref{soltj}) depends on the squared
tangent of $\sqrt{3 e_1/8}\, t$ which implies the period
\begin{equation}\label{P23C}
    P_j = 2 \pi \sqrt{\frac{2}{3 e_1(E_j)}}, \qquad j=2,3,
\end{equation}
where $e_1(E_j) = \sqrt[3]{g_3}= \sqrt{g_2/2}$. The same result can be obtained by taking the limit of (\ref{perin}) at $k_n \rightarrow 0$.

\subsubsection{Cassini state $C_4$ and homoclinic orbits $\Gamma_1$, $\Gamma_2$}

The energy of the unstable Cassini state $C_4$ can be evaluated from Eq.~(\ref{EofZ}) with $j=4$, and $Z^\ast_4$ given by
(\ref{Cas1234}). However, we need it not for the Cassini state itself, but rather to describe the motion on homoclinic orbits having the
energy $E_4$, which are $\Gamma_1$ and $\Gamma_2$. To this end, we will use the Safford form (\ref{solSaf}) with the reference points
given by Eq.~(\ref{z14q:f}), namely: $Z_4$ for $\Gamma_1$, and $Z_1$ for $\Gamma_2$.

Since $g_3<0$, the reduction (\ref{red3}) leads to
 \begin{eqnarray}
   X & = & \frac{\tilde{A} \sqrt{3 e_{12}}  \sinh(2 \tilde{u})}{\left(1+\tilde{B} \, \sinh^2\tilde{u} \right)^2},  \nonumber \\
   Y & = & Y_j + \frac{Z^2-Z_j^2}{2 a } = Y_j -  \frac{ 2 \tilde{A} \, \sinh^2\tilde{u} }{1+\tilde{B} \, \sinh^2\tilde{u} }
   \left( Z_j - \frac{ a \tilde{A} \, \sinh^2\tilde{u} }{1+\tilde{B} \, \sinh^2\tilde{u} }\right), \\
   Z &=& Z_j - \frac{ 2 a \tilde{A} \, \sinh^2\tilde{u} }{1+\tilde{B} \, \sinh^2\tilde{u} }, \nonumber
 \end{eqnarray}
 with the coefficients
 \begin{equation}
  \tilde{A} = \frac{ a b + Y_j Z_j}{3 e_{12}},
  \quad \tilde{B} =  \frac{e_{12}+ \frac{1}{3} \left(a Y_j + Z_j^2 \right) }{3 e_{12}},
 \end{equation}
where $e_{12}$ is given by Eq.~(\ref{e12e3}). The argument
 \begin{equation}
 \tilde{u} = \sqrt{3 e_{12}} ( \tau_0 + \phi_j),
 \end{equation}
 whose phase with respect to $Z_j$ at the initial epoch $t_0$ is given by
\begin{equation}
   \phi_j  =   \frac{\sgn{(X_0 \, \tilde{A})}}{\sqrt{3 e_{12}}} \,  \mathrm{arsinh}  \sqrt{\frac{Z_j-Z_0}{2 a \tilde{A} - \tilde{B} (Z_j-Z_0)}}.
\end{equation}
Note that the time rate of argument $\tilde{u}$ is the same at both the separatrices, and, as expected the solution tends to the Cassini state $C_4$
asymptotically at $t \rightarrow \pm \infty$.

\subsubsection{Cassini state $C_1$ and special orbit $\Gamma_3$}

The Cassini state $C_1$, being a stable equilibrium with $g_3>0$, is characterized by the period of small oscillations
given directly by the formula (\ref{P23C}) with the energy $E_1$ evaluated at $Y^\ast_1$, $Z^\ast_1$ deduced from Eq.~(\ref{Cas1234}).
What makes the difference, compared to $C_2$ or $C_3$, is the presence of another trajectory having the energy $E_1$ -- the special curve
$\Gamma_3$.

The motion along $\Gamma_3$ can be described using the Safford form solution (\ref{solSaf}) with respect to the turning points
$Z_1$ or $Z_2$, given by Eq.~(\ref{z12q:f}). Then, performing the reduction (\ref{red1}),
we obtain the solution in terms of trigonometric functions
 \begin{eqnarray}
   X & = & \frac{\bar{A} \gamma_3 \sin{2 \bar{u}}}{\left(1+ \bar{B} \sin^2\bar{u}\right)^2}, \nonumber \\
   Y & = & Y_j + \frac{Z^2-Z_j^2}{2 a } = Y_j - \frac{ \bar{A} \sin^2\bar{u}}{1+ \bar{B} \sin^2\bar{u}}
   \left( Z_j - \frac{a \bar{A} \sin^2\bar{u}}{2 \left(1+ \bar{B} \sin^2\bar{u} \right)}\right), \\
   Z &=& Z_j - \frac{a \bar{A} \sin^2\bar{u}}{1+ \bar{B} \sin^2\bar{u}}, \nonumber
 \end{eqnarray}
 where
 \begin{equation}
  \bar{A} = \frac{4 (a b + Y_j Z_j)}{3 e_1}, \quad \bar{B} = \frac{\frac{2}{3} \left(a Y_j + Z_j^2 \right) -e_1}{3 e_1}.
 \end{equation}
The argument $\bar{u}$ is
\begin{equation}
 \bar{u} = \gamma_3 (\tau_0 + \phi_j),
\end{equation}
where $\gamma_3 = \sqrt{3 e_1/2}$, as in Eq.~(\ref{ukg}) for $e_3= -e_1/2 $, hence the time period
of the solution is the same as that of small oscillations around $C_1$, i.e. $P_1$ given by Eq.~(\ref{P23C}).
The phase $\phi_j$ can be computed from
\begin{equation}
 \gamma_3 \phi_j = \sgn{(\bar{A} X_0)}\,\arcsin \sqrt{\frac{Z_j-Z_0}{a \bar{A} - (Z_j-Z_0) \bar{B}}}.
\end{equation}

The above solution can be obtained either from (\ref{solD14}) with $k_p=0$, and $e_3 =  -e_1/2 $,
or -- in a different form -- from (\ref{solD23}) with $k_n=0$, and $g_2=3 e_1^2$.

\subsubsection{Cassini state $C_{14}$ and special orbit $\Gamma_{23}$}

The most degenerate case occurs in type III, where $C_{14}$ is the cusp (parabolic) equilibrium with energy $E_{14}$ given by Eq.~(\ref{E14}).
The homoclinic curve $\Gamma_{23}$ with this energy can be parameterized using rational functions of time, as indicated by the reduction formula
(\ref{red0}). For the sake of using the Safford form, we introduce
\begin{equation}\label{u23}
    u_{23} = t-t_0 + \tau_{23}, \qquad \tau_{23} = t_0-t_\mathrm{m},
\end{equation}
where $\tau_{23}$ is the time interval between the initial epoch $t_0$ and the epoch of crossing the reference turning point
with coordinates $Y_\mathrm{m} = y_\mathrm{m}+a$, and $Z_\mathrm{m} =  z_\mathrm{m}-b$, as given by Eq.~(\ref{zmym3}).
Then, with $\beta = \sqrt[3]{a^4 b^2}$,
\begin{eqnarray}
  X  &=&  - \frac{8 a b u_{23}}{\left(1+\beta u_{23}^2 \right)^2}, \\
  Y &=&  Y_\mathrm{m} + \frac{4 a b u_{23}^2}{1+\beta u_{23}^2 } \left(Z_m + \frac{2 a^2 b u_{23}^2}{1+\beta u_{23}^2 } \right), \\
 Z &=&  Z_\mathrm{m} + \frac{4 a^2 b u_{23}^2}{1+\beta u_{23}^2 }.
\end{eqnarray}
The time offset $\tau_{23}$ is given by
\begin{equation}\label{tau23}
    \tau_{23} = - \sgn{X_0} \sqrt{\frac{Z_0-Z_\mathrm{m}}{4 a^2 b - \beta \left(Z_0-Z_\mathrm{m}\right)}}.
\end{equation}

\section{Conclusions}
\label{concl}
It is common to describe the integral curves of the Colombo top problem as an intersection of a parabolic cylinder and a
unit sphere, the latter being centered at the origin of the $x,y,z$ coordinate system. We have proposed a new point of view:
the curves can be tracked along the intersection of two out of the three invariant surfaces: a parabolic cylinder, a sphere, and a paraboloid of revolution.
This thread has been merely signaled, but it can be of possible interest when designing geometric integrators for the numerical treatment of the
Colombo top motion.
Moreover, using the shifted coordinates $X,Y,Z$, one introduces the symmetry to the parabolic cylinder on the paraboloid (at the expense of having an off-centered sphere)
which does simplify a number of expressions given in this work.

When partitioning the phase space of the Colombo top problem, we have completed the landscape, well known from earlier works, with an interesting but
hitherto overlooked feature: the trajectory $\Gamma_3$ which is unique by being periodic, yet expressible in terms of elementary functions of time.
Its presence calls for the distinction of $\mathrm{D}_3$ and $\mathrm{D}_4$ domains even though qualitatively they look  similar.
It also adds to a better understanding of the parametric bifurcation associated with the passage from type II, to type IV.

The analytical expressions for the turning points of the Colombo top trajectories as functions of energy, given in Sect.~\ref{inters}, had not been reported so far.
The expressions for the location of the Cassini states from Sect.~\ref{Cassi} depend only on parameters $a,b$. Up to some rearrangement of terms,
they are similar to those of \citet{SaLaBo:2019} in type II or III.
For type IV, when $a^\frac{2}{3}+b^\frac{2}{3} < 1$, the Cardano form provided by \citet{SaLaBo:2019} is
formally correct, but it gives real values only as the sums of two complex conjugates (\textit{casus irreducibilis} of the resolvent cubic).
In the present work we have preferred to use expressions based on the purely real trigonometric form whenever the quartic has a positive discriminant.

The differential equation for $\dot{Z}$, with its right-hand side proportional to the square root of the degree 4 polynomial, is not a novelty in celestial mechanics.
The same form pops up while discussing the Second Fundamental Model of resonance \citep{HenLem:1983}. Its solution in terms of the Weierstrass
elliptic function has always been given either in the simplified form of Eq.~(\ref{Bia:ZW}), as in \citep{FerMel:2007}, or in
the Biermann-Weierstrass form \citep{NesVok:2016}. We have taken an opportunity to recall other possibilities (Safford and Mordell forms)
than can be of use in other applications as well.

We hope that the results of the present work will facilitate the study of perturbed Colombo top problems. They should be useful either as the kernel
of analytical perturbation procedures, or as a building block of numerical integrators based upon composition methods.

\begin{acknowledgements}
The work of DV was funded by the Czech Science Foundation (grant 18-06083S).
\end{acknowledgements}
~\\
\noindent
\textbf{Compliance with ethical standards}\\
~\\
\textbf{Conflict of interest}{ The authors S. Breiter, J. Haponiak and D. Vokrouhlick\'y declare that they have no conflict of interest.
The article is partially based upon the doctoral thesis prepared by J. Haponiak, but it includes the results obtained independently by the co-authors.}


\appendix
\section{The roots of $W(Z)=0$}

\label{Ap:1}
\subsection{Factorization}
\label{fact:A}

In order to find the zeroes of the $W(Z)$, let us first write it in the `classical' polynomial form
\begin{equation} \label{Wagen}
W(Z) = a_0 Z^4 + 4 a_1 Z^3 + 6 a_2 Z^2 + 4 a_3 Z + a_4,
\end{equation}
where, according to Eq.~(\ref{Wgen}) the coefficients are
\begin{equation} \label{acoef}
a_0 = - 1, \quad a_1 = 0, \quad a_2 = - \frac{2 E}{3}, \quad a_3 = - 2 a^2 b, \quad
a_4 = - 4 E^2 + 8 a^2 E - 4 a^2 \left(a^2+b^2-1\right).
\end{equation}
Note the absence of the cubic term ($a_1=0$), meaning that equation $W(Z)=0$ is already in the reduced form.

Solving equation $W(Z)=0$, we essentially follow a simplified and slightly reformulated procedure of \citet{Neum:65}.
In particular, the absence of the third power of $Z$ allows the factorization
\begin{equation}
\label{a:fact}
W(Z) = a_0\, W_+(Z) \, W_-(Z) = a_0\, \left( Z^2 + 2 \sqrt{\xi}\, Z + h_+ \right) \left( Z^2 - 2 \sqrt{\xi}\, Z + h_- \right),
\end{equation}
with only three parameters ($\xi$, $h_+$, $h_-$), and three conditions resulting from equating the coefficients of $W$ and $W_+ W_-$:
\begin{equation}\label{a:conc}
    h_- h_+ = - a_4, \qquad h_- - h_+  = \frac{4 a^2 b}{\sqrt{\xi}}, \qquad h_- +h_+  = 4 \left( E  + \xi \right).
\end{equation}
The last two equations are easily solved for $h_-$ and $h_+$
\begin{equation}\label{a:h12a}
    h_\pm = 2 \left( E + \xi \mp \frac{a^2 b}{\sqrt{\xi}} \right),
\end{equation}
so the first of Eqs.~(\ref{a:conc}), after the substitution of Eq.~(\ref{a:h12a}), is actually the resolvent cubic equation
\begin{equation}\label{a:res}
    \xi^3 + 2 E \xi^2 + a^2 \left( 3 \rho  + 2 E \right) \xi - a^4 b^2 = 0,
\end{equation}
where $3\rho= 1-a^2-b^2$, according to Eq.~(\ref{alfa}).
The Descartes rule guarantees that (for nonzero $a$ and $b$) the resolvent has at least one positive real root to be used in factorization (\ref{a:fact}).

Before we proceed to solving the resolvent, let us make some important remarks. The four roots of the quartic equation $W(Z)=0$ come in two pairs of the roots of
$W_+(Z)=0$ and $W_-(Z)=0$, i.e.
\begin{eqnarray}
    W_+(Z)=0:&~~& Z_1 = - \sqrt{\xi} - \sqrt{\xi-h_+}, \qquad Z_2 =  - \sqrt{\xi} + \sqrt{\xi-h_+},  \nonumber \\
    \label{Z1234}
    W_-(Z)=0:&~~& Z_3 = \phantom{-} \sqrt{\xi} - \sqrt{\xi-h_-}, \qquad Z_4 =  \phantom{-}  \sqrt{\xi} + \sqrt{\xi-h_-},
\end{eqnarray}
Let $\xi_1$ be the only, or the greatest positive root of the resolvent (\ref{a:res}). Then, from the Vieta's formulas, we find for the remaining two roots
$\xi_2+\xi_3 = - 2 E - \xi_1$, and $\xi_2 \xi_3 = a^4 b^2 \xi_1^{-1}$, which allows to see that
\begin{equation} \label{X23h}
  \sqrt{\xi_2} \pm \sqrt{\xi_3}= \sqrt{\xi_2+\xi_3 \pm 2 \sqrt{\xi_2 \xi_3}} =\sqrt{ -\xi_1-2 E \pm \frac{ 2 a^2 b}{\sqrt{ \xi_1}}} = \sqrt{\xi_1 - h_\pm}.
\end{equation}
This leads to the Euler form of the solution
\begin{eqnarray}
    & & Z_1 = - \sqrt{\xi_1} - \sqrt{\xi_2} -\sqrt{\xi_3}, \quad Z_2 = - \sqrt{\xi_1} + \sqrt{\xi_2} + \sqrt{\xi_3}, \nonumber \\
   & &  Z_3 = \phantom{-} \sqrt{\xi_1} - \sqrt{\xi_2} + \sqrt{\xi_3}, \quad Z_4 = \phantom{-} \sqrt{\xi_1} + \sqrt{\xi_2}- \sqrt{\xi_3}. \label{Z1234E}
\end{eqnarray}

Assuming for the real roots $0 < \xi_3 \leq \xi_2 \leq \xi_1$, we guarantee a number of properties like
the ordering $Z_1 \leq Z_2 \leq Z_3 \leq Z_4$, the fact that a given trajectory contains only a pair $(Z_1,Z_2)$ or $(Z_3,Z_4)$, and that
if $\xi_2,\xi_3$ are complex conjugates, then $Z_1$ and $Z_2$ remain real, whereas $Z_3$ and $Z_4$ become complex.

\subsection{Weierstrass resolvent and its roots}
\label{Ap:Wr}

The cubic resolvent equation (\ref{a:res}) can be brought to a reduced form without the square term in a number of ways. We choose the substitution
based upon the seminvariant \citep{Jans:2011}
\begin{equation}\label{P2def}
    P_2 = a_0 a_2 - a_1^2 = \frac{2}{3} E,
\end{equation}
with
\begin{equation}\label{xitos}
    \xi =  s - P_2  =   s - \frac{2}{3} E.
\end{equation}
Applying it to Eq.~(\ref{a:res}), and multiplying both sides by 4, we obtain
\begin{equation}\label{Sdef}
    S(s) = 4 s^3 - g_2 s - g_3 = 0.
\end{equation}
The cubic polynomial $S(s)$ plays a special role in the theory of the Weierstrass elliptic functions, thus let us call it the Weierstrass resolvent.
The symbols $g_2$ and $g_3$ that appear in Eq.~(\ref{Sdef}) are the two, algebraically independent, basis invariants of the quartic $W(Z)$:
\begin{itemize}
  \item the apolar invariant of degree 2
  \begin{equation}\label{g2}
    g_2 = a_0 a_4 + 3 a_2^2 - 4 a_1 a_3 = 12 \left[ \left(\frac{2 E}{3}\right)^2 - a^2 \left(\frac{2 E}{3}\right) - a^2 \rho \right],
  \end{equation}
  \item the Hankel determinant of degree 3
  \begin{equation}\label{g3}
    g_3 = \left|
            \begin{array}{ccc}
              a_0 & a_1 & a_2 \\
              a_1 & a_2 & a_3 \\
              a_2 & a_3 & a_4 \\
            \end{array}
          \right| =  -8 \left(\frac{2 E}{3}\right)^3 + 12 a^2 \left(\frac{2 E}{3}\right)^2 +12 a^2 \rho \left(\frac{2 E}{3}\right)
          + 4 a^4 b^2,
  \end{equation}
 known also as a catalecticant \citep{Jans:2011}.
\end{itemize}

Notably, both the discriminants: $\Delta_4$ of the quartic $W(Z)$, and $\Delta_3$ of the cubic $S(s)$ are not only expressible in terms of $g_2$ and $g_3$,
but they are equal up to a constant factor. If
\begin{equation}\label{Delta}
    \Delta = \left(\frac{g_2}{3}\right)^3- g_3^2,
\end{equation}
then $\Delta_4 = 16 \Delta_3 = 4^4\,3^3 \Delta$.

Let the roots of the Weierstrass resolvent equation $S(s)=0$ be $s=e_1$, $s=e_2$, and $s=e_3$. By the Vieta's formulae, they satisfy
\begin{equation} \label{e:V}
e_1+e_2+e_3=0, \qquad e_1e_2+e_1 e_3 + e_2 e_3 = - \frac{g_2}{4}, \qquad e_1 e_2 e_3 = \frac{g_3}{4},
\end{equation}
and by the definition of the scaled discriminant (\ref{Delta})
\begin{equation} \label{e:D}
\left(e_1-e_2\right)^2\left(e_1-e_3\right)^2\left(e_2-e_3\right)^2= \frac{27\Delta}{16}.
\end{equation}
Introducing auxiliary quantities $\beta$ and $\phi$, such that
\begin{equation}\label{betfi}
    g_2 = 3 \beta^2, \qquad g_3 = \beta^3 \cos{\phi}, \qquad \Delta =  \beta^6 \sin^2\phi,
\end{equation}
hence
\begin{equation}\label{betfig}
    \beta = \sqrt{\frac{g_2}{3}}, \qquad \cos \phi =   g_3 \left(\frac{3}{g_2}\right)^\frac{3}{2},
\end{equation}
we can establish the universal formula for the roots \citep{Briz:2015}
\begin{eqnarray}
    e_1 & = &  \beta \cos{\frac{\phi}{3}},  \nonumber \\
    e_2 & = &  \beta  \cos{\frac{\phi-2\pi}{3}} = - \frac{e_1}{2} + \frac{\sqrt{3}}{2} \beta \sin{\frac{\phi}{3}}
    = - \frac{e_1}{2} + \frac{\sqrt{g_2-3 e_1^2}}{2},  \label{ejBr} \\
    e_3 & = &  \beta \cos{\frac{\phi+2\pi}{3}} =  - \frac{e_1}{2}- \frac{\sqrt{3}}{2} \beta \sin{\frac{\phi}{3}}
    =   - \frac{e_1}{2} - \frac{\sqrt{g_2-3 e_1^2}}{2}. \nonumber
\end{eqnarray}

If  $\Delta > 0$ (hence $g_2>0$) there are three simple real roots $e_3 < e_2 < e_1$ given directly by Eqs.~(\ref{ejBr}).
When $\Delta < 0$, there is one real root $e_1$ and two complex ones, with $e_2 = \overline{e}_3$. Equations (\ref{betfi})
and (\ref{ejBr}) remain valid in principle, but they involve complex quantities and require distinguishing the sign of $g_2$. In these circumstances
it is more convenient to use the Cardano form for the real root\footnote{In this approach, $\sqrt[3]{x}$ of a real argument $x$ is used as a real-valued function for $x<0$, i.e. $\sqrt[3]{-1} = -1$.}
\begin{equation}\label{e1:dm}
    e_1  = \frac{1}{2} \left( \sqrt[3]{g_3 + \sqrt{ -\Delta }} + \sqrt[3]{g_3 - \sqrt{ -\Delta }}\right).
\end{equation}
and
\begin{equation}\label{e23:dm}
    e_2 = - \frac{e_1- \mathrm{i} e_\mathrm{c}}{2}, \qquad e_3 = \bar{e}_2 = - \frac{e_1 + \mathrm{i} e_\mathrm{c}}{2},
\end{equation}
with
\begin{equation}\label{ec:def}
    e_\mathrm{c} =  \sqrt{3 e_1^2- g_2}  = \frac{\sqrt{3}}{2} \left( \sqrt[3]{g_3 + \sqrt{ -\Delta }} - \sqrt[3]{g_3 - \sqrt{ -\Delta }} \right),
\end{equation}
for the complex roots.

Finally, the degeneracy $\Delta=0$ implies real roots: one simple and one double, or one triple root.
The former case requires $g_3 \neq 0$ and $g_2 > 0$; then, according to the sign of $g_3$, either
  \begin{equation}\label{e1e23}
    e_1 =  \sqrt[3]{g_3}=\sqrt{\frac{g_2}{3}}, \qquad e_2=e_3=e_{23} = - \frac{e_1}{2}, \mbox{~~~for $g_3 > 0$}
  \end{equation}
or
  \begin{equation}\label{e12e3}
    e_3 = \sqrt[3]{g_3}= -\sqrt{\frac{g_2}{3}}, \qquad e_1 = e_2 = e_{12} =  - \frac{e_3}{2}, \mbox{~~~for $g_3 < 0$}.
  \end{equation}
The ordering of roots in (\ref{e12e3}) is exceptional ($e_3$ is the greatest), but helps to maintain a coherent notation in
further applications.

The triple root $e_{123}=0$ may appear only for $g_2=g_3=0$, which is possible only when $a^\frac{2}{3} + b^\frac{2}{3} = 1$.

\subsection{The roots $Z_j$}

Although the roots of $W(Z)$ are to be expressed in terms of the roots of $S(s)$, we need to include in the discussion not only the invariants
$\Delta$, $g_2$, and $g_3$, but also seminvariants $P_2$ and
\begin{equation} \label{Q2:def}
  Q_2 =   2 a_0^2 a_3 -  6  a_0 a_1 a_2  + 4 a_1^3 = - 4 a^2 b.
\end{equation}
This is due to the fact that although formally it is enough to substitute
\begin{equation}\label{subr}
    \xi_j = e_j - \frac{2}{3} E = e_j - P_2,
\end{equation}
into (\ref{Z1234E}), the signs of $\xi_j$ play a significant role in determining which of the roots are real and which are complex,
and there are various ways to create multiple roots.

In the following discussion we will refer to the Theorem 9.3 of \citet{Jans:2011}, adjusted to the different scaling of our invariants and seminvariant
 (namely, his $P= 48 P_2$, $Q=16 Q_2$, $J= 432 g_3$, and $I=12 g_2$).

\subsubsection{Four simple real roots ($\Delta >0$ and all $\xi_j \geq 0$)}

If the four real roots exist, they take the form (\ref{Z1234E}) with $\xi_j$ defined in Eq.~(\ref{subr}) and $e_j$ as in (\ref{ejBr}).
This requires not only $\Delta > 0$ to have three simple real roots $e_j$, but also that $\xi_j \geq 0$ for each $j \in \{1,2,3\}$. The latter
is secured by $P_2 < 0$ and $12 P_2^2 - a_0^2 g_2 \geq 0$ \citep{Jans:2011}. Substituting (\ref{P2def}), (\ref{g2}) and (\ref{acoef}), we obtain
\begin{equation}\label{cnd:4r}
     \Delta > 0,  \mbox{~~and~~} - \rho \leq \frac{2E}{3} < 0,
\end{equation}
as the  condition for the real quadruple $Z_1 < Z_2 < Z_3 < Z_4$. If $\Delta$ is positive, but the second condition in (\ref{cnd:4r}) is not fulfilled,
there are no real roots, and $Z_j$ form two distinct pairs of complex conjugate numbers.

\subsubsection{Two simple real roots ($\Delta < 0$)}

When $e_2$ and $e_3$ are complex, the pair $(Z_3,Z_4)$ is complex, whereas $(Z_1,Z_2)$ in the formula (\ref{Z1234E}) formally remain real-valued,
yet only by canceling the imaginary parts. Using $\xi_1 = e_1 - P_2$, with $e_1$ given by Eq.~(\ref{e1:dm}), we can obtain $Z_1$ and $Z_2$ directly
from Eqs.~(\ref{Z1234}) and (\ref{a:h12a}). Alternatively, we can find the expressions for $\sqrt{\xi_2} \pm \sqrt{\xi_3}$, which results in
\begin{equation}\label{Z12:dm}
   Z_1 = - \sqrt{\xi_1} - \sqrt{2|\xi_2|  - \xi_1 - 3 P_2}, \qquad   Z_2 = - \sqrt{\xi_1} + \sqrt{2 |\xi_2| - \xi_1 - 3 P_2},
\end{equation}
and
\begin{equation}\label{Z34:dm}
   Z_3 =  \sqrt{\xi_1} - \mathrm{i}\, \sqrt{ 2 |\xi_2| + \xi_1 + 3 P_2}, \qquad   Z_4 = \overline{Z}_3,
\end{equation}
where
\begin{equation}\label{lab}
 \xi_1 = e_1 - P_2, \qquad  2 |\xi_2| = 2 |\xi_3| = |e_1 + 2 P_2 + \mathrm{i} e_\mathrm{c}| =  \sqrt{\left(e_1+ 2 P_2\right)^2 + e_\mathrm{c}^2},
\end{equation}
with $e_1$  given by Eq.~(\ref{e1:dm}).

\subsubsection{Multiple roots ($\Delta = 0$)}

The statement $\Delta=0$ means only that at least one of the roots is at least a double root. Further distinction is based upon the signs and values of
$g_2$, $g_3$ and $P_2$. Let us inspect five possibilities involving multiple real roots from the Theorem 9.3 of \citet{Jans:2011}.

A quadruple real root is not possible, because it requires $P_2=g_2=g_3=0$, whereas substituting $E=0$ we obtain $g_3= 4 a^4 b^2 \neq 0$.
Two real double roots are also impossible, because they require (among other conditions) that  $Q_2 = 0$, which is not the case.
The remaining three cases are the following.
\begin{enumerate}
\item A triple real root and one single real root appear when $g_2=g_3=0$, and $P_2 < 0$.
Taking the resultant of $g_2$ and $g_3$ considered as the polynomials in $E$, one finds that both the invariants admit the common root
if
\begin{equation}
    a^\frac{2}{3} + b^\frac{2}{3} = 1,
\end{equation}
the relation well known from \citet{HenMur:1987}. With this constraint, $g_2=0$ can be solved to give a unique negative root
\begin{equation}\label{Etr}
    E = - \frac{3}{2} \left( a^2 b\right)^\frac{2}{3}.
\end{equation}
According to the statement below Eq.~(\ref{e12e3}), $g_2=g_3=0$ refers to the triple root $e_{123}=0$, hence, with
$\xi_1=\xi_2=\xi_3= -\frac{2}{3}E$, we obtain
\begin{equation}\label{Z1Z234}
    Z_1 = - 3 \sqrt[3]{a^2 b}, \qquad Z_{234} = \sqrt[3]{a^2 b},
\end{equation}
where $Z_1$ is the single, and $Z_{234}$ is the triple root.
\item Two simple real roots $Z_1$, $Z_2$ and double real root $Z_{34}$ appear when $g_2 > 0$, $P_2<0$, and  $12 P_2^2-g_2 > 0$. So, if
$E$ is a real root of $\Delta = 0$ in the interval
\begin{equation}\label{con:m1}
   -\rho  <  \frac{2E}{3} < 0,
\end{equation}
then either
\begin{equation}\label{z:m1}
    Z_1 = - \sqrt{\xi_1} - 2 \sqrt{\xi_{23}}, \qquad Z_2 = - \sqrt{\xi_1} + 2 \sqrt{\xi_{23}}, \qquad Z_{34} = \sqrt{\xi_1},
    \qquad \mbox{for~} g_3 > 0,
\end{equation}
or
\begin{equation}\label{z:m1a}
    Z_1 = - \sqrt{\xi_3} - 2 \sqrt{\xi_{12}}, \qquad Z_{23} = \sqrt{\xi_3}, \qquad Z_{4} = - \sqrt{\xi_3} + 2 \sqrt{\xi_{12}}, \qquad \mbox{for~} g_3 < 0,
\end{equation}
where $\xi_i = e_i - P_2$, and $\xi_{ij}=e_{ij}-P_2$, with the Weierstrass resolvent roots given by Eq.~(\ref{e1e23}) or
(\ref{e12e3}), according to the sign of $g_3$.
\item If the energy $E$ is a real root of $\Delta=0$ outside the interval (\ref{con:m1}), i.e.
\begin{equation}\label{con:m2}
   \frac{2E}{3} < -\rho, \mbox{~~or~~} \left( E > 0, \mbox{~~and~~} \frac{2E}{3} \neq  -\rho \right),
\end{equation}
then a double real root $Z_{12}$ is accompanied by two simple complex roots $Z_3$ and $Z_4$. This case appears when $e_\mathrm{c}=0$
in Eq.~(\ref{e23:dm}). Accordingly,
\begin{equation}\label{Z12}
    Z_{12} = - \sqrt{ \xi_1 } = - \sqrt{e_1-P_2},
\end{equation}
where $e_1$ is given by Eq.~(\ref{e1e23}).
\end{enumerate}

\section{Cassini states coordinates $z^\ast_j$}
\label{Ap:Cas}

The left-hand side of the  quartic equation $w(z-b)=0$ is the polynomial
\begin{equation}\label{wzb}
    w(z-b) = a'_0 z^4 + 4 a'_1 z^3 + 6 a'_2 z^2 + 4 a'_3 z + a'_4,
\end{equation}
where
\begin{equation} \label{apcoef}
a'_0 =  -1, \quad a'_1 =  \frac{b}{2}, \quad a'_2 = \frac{1-a^2-b^2}{6} =  \frac{\rho}{2} , \quad a'_3 = - \frac{b}{2}, \quad
a'_4 = b^2.
\end{equation}

Evaluating the invariants and seminvariants from the primed coefficients, we find
\begin{equation}\label{invp}
    g'_2 = \frac{3\rho^2}{4} \geq 0, \qquad g'_3 =   -\left(\frac{\rho}{2}\right)^3 + \frac{a^2 b^2}{4},
    \qquad P'_2 = - \frac{\rho}{2} - \frac{b^2}{4},\qquad Q'_2 = - \frac{ \left(1 + a^2\right) b}{2} < 0,
\end{equation}
ant the scaled discriminant is
\begin{equation}\label{deltap}
    \Delta' = \left(\frac{g'_2}{3}\right)^3 - \left(g'_3\right)^2 =  \frac{a^2 b^2}{16}\left( \rho^3 -  a^2 b^2 \right).
\end{equation}

The transformation
\begin{equation}\label{zr}
   z = z_\mathrm{r}+\frac{b}{2},
\end{equation}
converts the equation $w(z-b)=0$ into $w(z_\mathrm{r}-b/2)=0$, which is free of the $z_\mathrm{r}^3$ term, and so is ready for the factorization
from Sect.~\ref{fact:A}. Fortunately, we do not need to know the coefficients of the equation in $z_\mathrm{r}$, because we require only the
invariants (\ref{invp}) which are conserved under the simple transformation (\ref{zr}). Thus, tracing backward the procedure from Appendix~\ref{Ap:1},
we start with solving the Weierstrass resolvent
\begin{equation}\label{wresp}
    4 s^3 - g'_2 s - g'_3 = 0,
\end{equation}
finding the roots $e'_1$, $e'_2$, and $e'_3$. These define
\begin{equation}\label{xip}
    \xi'_j =    e'_j - P'_2 = e'_j + \frac{\rho}{2} + \frac{b^2}{4},
\end{equation}
as in Eqs.~(\ref{xitos}) and (\ref{subr}). Finally, four roots $z_{\mathrm{r},i}$ are given by Eq.~(\ref{Z1234E})
 with $Z_i \rightarrow z_{\mathrm{r},i}, \xi_j \rightarrow \xi'_j$, and then $z^\ast_i = z_{\mathrm{r},i} + (b/2)$.
Each real solution $z^\ast_i$ is the $z$ coordinate of some Cassini state.

The number of real roots of $w(z-b)=0$ depends on the sign of $\Delta'$.
\begin{enumerate}
\item $\Delta' > 0$ is equivalent to $a^\frac{2}{3}+b^\frac{2}{3} < 1$. Three real roots $e'_1>e'_2>e'_3$ of the resolvent (\ref{wresp})
are, by analogy with (\ref{ejBr}),
\begin{eqnarray}
    e'_1 & = &  \frac{\rho}{2} \cos{\frac{\phi'}{3}},  \nonumber \\
    e'_2 & = &  \frac{\rho}{2}  \cos{\frac{\phi'-2\pi}{3}}
    = - \frac{e'_1}{2} + \frac{\sqrt{g'_2-3 (e'_1)^2}}{2},  \label{ejBrp} \\
    e'_3 & = &  \frac{\rho}{2} \cos{\frac{\phi'+2\pi}{3}}
    =   - \frac{e'_1}{2} - \frac{\sqrt{g'_2-3 ( e'_1)^2}}{2}, \nonumber
\end{eqnarray}
where
\begin{equation}\label{cfp}
    \phi' = \arccos\left(\frac{2 a^2 b^2}{\rho^3}-1\right) = \pi - 2 \arcsin\left( \frac{a b}{\rho^\frac{3}{2}}\right).
\end{equation}

They always define four real roots $z^\ast_i$, because all  $\xi'_j$ are positive. Indeed
\begin{eqnarray}
    \xi'_1 & = &   \frac{b^2}{4} + \frac{\rho}{2}\left(1+\cos{\frac{\phi'}{3}}\right)  =\frac{b^2}{4}+ \rho \cos^2\frac{\phi'}{6} > 0, \nonumber \\
    \xi'_2  & = &  \frac{b^2}{4} + \frac{\rho}{2}\left(1+\cos{\frac{\phi'-2\pi}{3}}\right) =
     \frac{b^2}{4}+ \rho \cos^2\frac{\phi'-2\pi}{6} > 0, \label{xip:1} \\
    \xi'_3  & = &  \frac{b^2}{4} + \frac{\rho}{2}\left(1+\cos{\frac{\phi'+2\pi}{3}}\right) =
     \frac{b^2}{4}+ \rho \cos^2\frac{\phi'+2\pi}{6} > 0. \nonumber
\end{eqnarray}
Unlike in eq.~(\ref{Z1234E}), we label the roots $z^\ast_i$ not according to their ordering in magnitude, but so that the subscript $i$ matches the the Cassini state label $C_i$ according to \citet{Colo:1966AJ}, it is
\begin{eqnarray}
    & & z^\ast_1 = \frac{b}{2} + \sqrt{\xi'_1} + \sqrt{\xi'_2} - \sqrt{\xi'_3}, \quad z^\ast_2 = \frac{b}{2} - \sqrt{\xi'_1} + \sqrt{\xi'_2} + \sqrt{\xi'_3}, \nonumber \\
   & &  z^\ast_3 = \frac{b}{2} - \sqrt{\xi'_1} - \sqrt{\xi'_2} - \sqrt{\xi'_3}, \quad z^\ast_4 = \frac{b}{2} + \sqrt{\xi'_1} - \sqrt{\xi'_2} + \sqrt{\xi'_3}, \label{z1234cas}
\end{eqnarray}
with $z^\ast_3 < z^\ast_2 < z^\ast_4 < z^\ast_1$, as expected.
\item $\Delta' < 0$ means $a^\frac{2}{3}+b^\frac{2}{3} > 1$. The Weierstrass resolvent has one real root $e'_1$ and two complex roots $e'_2 = \overline{e'}_3$
given by Eqs.~(\ref{e1:dm}), (\ref{e23:dm}), and (\ref{ec:def}) with $g'_2$, $g'_3$, and $\Delta'$.
Accordingly, we obtain two real roots $z^\ast_i$ -- the ones involving $\sqrt{\xi'_2}+\sqrt{\xi'_3}$, where the imaginary part cancels out. Adapting the expressions
(\ref{Z12:dm}) and (\ref{lab}), and adjusting the subscripts of $z^\ast_i$ to the Cassini states $C_3$ and $C_2$, we obtain two real roots of $w(z-b)=0$ as
\begin{equation}\label{z32c}
   z^\ast_3 = \frac{b}{2} - \sqrt{\xi'_1} - \sqrt{2|\xi'_2|  - \xi'_1 - 3 P'_2}, \qquad   z^\ast_2 =\frac{b}{2} - \sqrt{\xi'_1} + \sqrt{2 |\xi'_2| - \xi'_1 - 3 P'_2},
\end{equation}
with $z^\ast_3 < z^\ast_2$. The final substitution is made in Sect.~\ref{Cassi}.
\item $\Delta'=0$, hence $a^\frac{2}{3}+b^\frac{2}{3} = 1$, implies one single and one double real root of the Weierstrass resolvent. A triple root is excluded, because
$g'_2 = \frac{3}{4}(ab)^\frac{4}{3} \neq 0$. Thus, observing that $g'_3 = a^2 b^2/8 > 0$, and $\rho = (ab)^\frac{2}{3}$,
\begin{equation}\label{e1p}
    e'_{1} = \frac{(ab)^\frac{2}{3}}{2}, \qquad e'_{23} = - \frac{ e'_{1}}{2},  \qquad \xi'_{1}= \frac{4 a^\frac{2}{3}b^\frac{2}{3} + b^2}{4}, \qquad
    \xi'_{23} = \frac{a^\frac{2}{3}b^\frac{2}{3} + b^2}{4},
\end{equation}
provide three Cassini states: two usual $C_3$, $C_2$, and one degenerate $C_{14}$, with
\begin{equation}\label{Casdeg}
   z^\ast_3 =  \frac{b}{2}- \sqrt{\xi'_1} - 2 \sqrt{\xi'_{23}},
    \qquad z^\ast_2 =  \frac{b}{2} - \sqrt{\xi'_1} + 2 \sqrt{\xi'_{23}}, \qquad  z^\ast_{14} = \frac{b}{2} + \sqrt{\xi'_1},
\end{equation}
listed in the ascending order.
\end{enumerate}

\end{document}